\documentclass[%
 reprint,
nofootinbib,
 amsmath,amssymb,
 aps,
]{revtex4-2}

\usepackage{graphicx}
\usepackage{dcolumn}
\usepackage{bm}
\usepackage[colorlinks=true,allcolors=blue]{hyperref}
\usepackage{makecell}
\usepackage{tabularx}
\usepackage{array}
\usepackage{nicefrac}
\newcolumntype{Y}{>{\centering\arraybackslash}X}
\renewcommand{\arraystretch}{1.2}
\usepackage{orcidlink}

\newcommand{\muKarcmin}{$\mu\text{K-arcmin}$}

\begin{document}

\preprint{APS/123-QED}

\title{The Atacama Cosmology Telescope: $B$-mode delensing with DR6 data and external tracers of large-scale structure}

\author{Emilie Hertig\,\orcidlink{0000-0001-9189-4035}~$^{1,2}$}\email{emh83@cam.ac.uk}
\author{Antón Baleato Lizancos\,\orcidlink{0000-0002-0232-6480}~$^{3,4,5}$}
\author{Frank J. Qu\,\orcidlink{0000-0001-7805-1068}~$^{6,7}$}
\author{J. Richard Bond\,\orcidlink{0000-0003-2358-9949}~$^{8}$}
\author{Erminia~Calabrese\,\orcidlink{0000-0003-0837-0068}~$^9$}
\author{Anthony Challinor\,\orcidlink{0000-0003-3479-7823}~$^{1,2,10}$}
\author{Mark J. Devlin\,\orcidlink{0000-0002-3169-9761}~$^{11}$}
\author{Jo Dunkley\,\orcidlink{0000-0002-7450-2586}~$^{12,13}$}
\author{Thibaut Louis\,\orcidlink{ 0000-0002-6849-4217}~$^{14}$}
\author{Mathew S. Madhavacheril\,\orcidlink{ 0000-0001-6740-5350}~$^{11}$}
\author{Toshiya Namikawa \orcidlink{0000-0003-3070-9240}~$^{9,15,2}$}
\author{Lyman A. Page~$^{12}$}
\author{Neelima~Sehgal\,\orcidlink{ 0000-0002-9674-4527}~$^{16}$}
\author{Blake Sherwin\,\orcidlink{0000-0002-4495-1356}~$^{10,2}$}
\author{Crist\'obal Sif\'on\,\orcidlink{0000-0002-8149-1352}~$^{17}$}
\author{Suzanne T. Staggs\,\orcidlink{ 0000-0002-7020-7301}~$^{12}$}
\author{Edward~J.~Wollack\,\orcidlink{0000-0002-7567-4451}~$^{18}$}

\affiliation{$^1$Institute of Astronomy, University of Cambridge, Madingley Road, Cambridge, CB3 0HA, UK}
\affiliation{$^2$Kavli Institute for Cosmology Cambridge, Madingley Road, Cambridge, CB3 0HA, UK}
\affiliation{$^3$Berkeley Center for Cosmological Physics, UC Berkeley, CA 94720, USA}
\affiliation{$^4$Department of Physics, UC Berkeley, CA 94720, USA}
\affiliation{$^5$Lawrence Berkeley National Laboratory, One Cyclotron Road, Berkeley, CA 94720, USA}
\affiliation{$^6$Kavli Institute for Particle Astrophysics and Cosmology, Stanford University, 452 Lomita Mall, Stanford, CA 94305, USA}
\affiliation{$^7$Department of Physics, Stanford University, 382 Via Pueblo Mall, Stanford, CA 94305, USA}
\affiliation{$^8$Canadian Institute for Theoretical Astrophysics, University of Toronto, 60 St. George Street, Toronto, ON, M5S 3H8, Canada}
\affiliation{$^9$School of Physics and Astronomy, Cardiff University, The Parade, Cardiff, Wales, UK CF24 3AA} 
\affiliation{$^{10}$DAMTP, University of Cambridge, Cambridge, CB3 0WA, UK}
\affiliation{$^{11}$Department of Physics and Astronomy, University of Pennsylvania, 209 South 33rd Street, Philadelphia, PA 19104, USA}
\affiliation{$^{12}$Joseph Henry Laboratories of Physics, Jadwin Hall, Princeton University, Princeton, NJ 08544, USA}
\affiliation{$^{13}$Department of Astrophysical Sciences, Peyton Hall, Princeton University, Princeton, NJ 08544, USA}
\affiliation{$^{14}$Université Paris-Saclay, CNRS/IN2P3, IJCLab, 91405 Orsay, France}
\affiliation{$^{15}$Center for Data-Driven Discovery, Kavli IPMU (WPI), UTIAS, The University of Tokyo, Kashiwa, 277-8583, Japan}
\affiliation{$^{16}$Physics and Astronomy Department, Stony Brook University, Stony Brook, NY USA 11794}
\affiliation{$^{17}$Instituto de F\'isica, Pontificia Universidad Cat\'olica de Valpara\'iso, Casilla 4059, Valpara\'iso, Chile}
\affiliation{$^{18}$NASA Goddard Space Flight Center, 8800 Greenbelt Road, Greenbelt, MD 20771, USA}

\collaboration{The ACT Collaboration}

\date{\today}

\begin{abstract} 
    Large-scale $B$-mode polarization of the cosmic microwave background (CMB) is a prime target for current and future experiments in search of primordial gravitational waves (PGW). With increasingly sensitive instruments being deployed, secondary $B$-modes induced by the weak gravitational lensing of CMB photons are becoming an important limitation and need to be removed, a process known as delensing. In this work, we combine internally reconstructed CMB lensing maps from the Atacama Cosmology Telescope (ACT) DR6 with galaxy samples from unWISE and a map of the cosmic infrared background (CIB) fluctuations from Planck to produce a well-correlated tracer of the CMB lensing field. Our co-added tracer, shown to be 55--85\% correlated with the true lensing convergence at multipoles $L \leq 2000$, is then convolved with ACT DR6 $E$-mode polarization to yield a template of the lensing $B$-modes.
    We assess its performance on a wide range of scales by using it to delens ACT DR6 and Planck $B$-modes over 23\% of the sky, removing around 39\% of the lensing power at $100\leq l \leq 1500$ and 47\% at $30 \leq l \leq 300$, respectively. Our template achieves the highest delensing efficiency to date and will be useful for the analysis of early polarization maps from the Simons Observatory (SO). We finally outline prospects for further improvements by including additional large-scale structure tracers from upcoming cosmological surveys.
\end{abstract}

\maketitle


\section{Introduction}

Decades of cosmological observations have revealed a remarkably simple picture of our Universe: spatially flat, statistically homogeneous and isotropic on large scales, and well-described by a six-parameter phenomenological model with a nearly-scale invariant sprectrum of primordial density fluctuations~\cite{planck_collab_cosmo_params_2020,calabrese_actdr6extended_2025,camphuis_SPT3G_2025}. Yet, this apparent simplicity hints at a deeper issue as it could not have been realized without precisely adjusted initial conditions. Inflation, a hypothetical period of exponential expansion within the first fraction of a second~\cite{Brout1978,Starobinsky1980,Kazanas1980,Sato1981,guth_inflationary_1981,AlbrechtSteinhardt1982,Linde1982,Linde1983}, provides an answer to this fine-tuning question as well as a direct connection between primordial quantum dynamics and the large-scale structure of the Universe at later times. Within this framework, scalar fluctuations of the field(s) driving inflation would have been stretched from microscopic to macroscopic scales and morphed into density variations as the origin of all cosmic structures. Similarly, tensor perturbations resulting from fluctuations of the spacetime metric or from the presence of gauge fields may have produced a detectable background of primordial gravitational waves (PGW)~\cite{starobinskii_spectrum_1979,mukhanov_quantum_1981,abbott_constraints_1984}. The search for PGW is one of the most crucial quests in modern cosmology, as a detection would indicate the energy scale of inflation and shed light on fundamental physics at the highest energies.

The imprints of primordial fluctuations are encoded in the cosmic microwave background (CMB)~\cite{ZaldarriagaSeljak1997_Analysis,kamionkowski_statistics_1997}, with PGW producing distinctive parity-odd $B$-mode polarization patterns on large scales~\cite{seljak_signature_1997,Kamionkowski1997_Probe}. While such $B$-modes have not been observed yet, measurements from WMAP~\cite{hinshaw_WMAP9_2013}, Planck~\cite{Planck2018_Inflation} and BICEP/Keck Array~\cite{bicepkeck_collaboration_bicep_2021} have set tight upper bounds on their amplitude parametrized by the tensor-to-scalar ratio $r$. Most recently, Ref.~\cite{bicepkeck_collaboration_bicep_2021} found $r<0.036$ at 95\% confidence for a pivot scale of 0.05~Mpc$^{-1}$ and a slightly lower value of $r<0.032$ was inferred by adding the Planck PR4 maps~\cite{tristram_improved_2022}; these results rule out some of the simplest classes of inflation models. Next-generation CMB experiments are expected to yield increasingly stringent constraints on $r$ in the coming decade, aiming to test popular models such as Starobinsky~\cite{Starobinsky1980} and Higgs inflation~\cite{bezrukov_standard_2008} decisively. For example, the Simons Observatory (SO)~\cite{the_simons_observatory_collaboration_simons_2019}, a ground-based instrument that has been collecting data since its first light in October 2023, is forecasted to achieve a precision level $\sigma(r=0)\lesssim0.003$ by the end of its initial five-year survey. Similar constraints are anticipated from the South Pole Observatory (SPO), the combination of the BICEP Array and the South Pole Telescope~\cite{cmbs4_revised_2025}. Uncertainties on $r$ should be further reduced as planned SO hardware extensions come online, and could also improve with proposed upgrades of SPO; $\sigma(r)$ is projected to fall below $10^{-3}$ with the LiteBIRD mission~\cite{litebird_collaboration_probing_2023} in the 2030s.

Unlike past experiments, which were mainly limited by noise, upcoming missions will be increasingly sensitive to other contaminating sources of $B$-mode polarization. Beside Galactic foregrounds, which can be distinguished from the CMB through multi-frequency observations~\cite{katayama_simple_2011,thorne_removal_2019,chluba_rethinking_2017}, one such source is the deflection of CMB photons by large-scale structure (LSS). This weak gravitational lensing effect results in a direction-dependent remapping of polarization anisotropies, causing part of the $E$-mode signal produced by scalar fluctuations to morph into $B$-modes~\cite{ZaldarriagaSeljak1998_Lensing,lewis_weak_2006,hanson_detection_2013}. Although lensing-induced $B$-modes are not expected to bias PGW measurements as their power spectrum is well-constrained and can be accounted for, their contribution to sample variance increases statistical uncertainties on $r$. On large scales, this effect is roughly equivalent to 5\,\muKarcmin\ white noise and is significant in comparison to the goal noise level of 2\,\muKarcmin for SO's initial five-year survey~\cite{the_simons_observatory_collaboration_simons_2019,namikawa_simons_2022}. Reducing this additional variance cannot be done simply by subtracting the average lensing $B$-mode power; it requires modelling and removing the specific $B$-mode pattern created in our sky by the actual realization of the lensing potential. This coherent removal process, known as \textit{delensing}, is becoming a necessity to reach the target precision of current and future PGW searches.

Using an empirical proxy of the CMB lensing convergence field $\kappa$, the lensing $B$-mode realization in our sky can be estimated non-perturbatively by remapping observed polarization maps to reverse the deflection. This technique was applied to delens Planck and ACT maps in~\cite{carron_internal_2017,planck_collaboration_planck_2020_lensing} and~\cite{han_atacama_2020}, respectively; it was also used in~\cite{bicepkeck_and_sptpol_collaborations_demonstration_2021}, where a delensing-related improvement in $\sigma(r)$ was demonstrated for the first time on real data. The alternative approach, which this paper focuses on, consists of convolving the $\kappa$ estimator with measured $E$-modes in harmonic space to obtain a gradient-order template of the lensing $B$-modes~\cite{hanson_detection_2013,manzotti_cmb_2017}. Besides having the advantage of a more straightforward implementation, this method was shown to be effectively optimal for our purposes as a result of cancellations between higher-order terms in the delensed $B$-mode power spectrum~\cite{lizancos_limitations_2021}. In this work, we construct the most accurate gradient-order lensing template based on existing datasets; our results will be particularly relevant to the delensing of early SO $B$-modes, and our methodology will be straightforward to extend in order to include high-resolution data from SO's Large Aperture Telescope (LAT) as it becomes available.

Our template uses $E$-modes and internal CMB lensing convergence maps from the Atacama Cosmology Telescope (ACT) data release 6 (DR6)~\cite{qu_2024}. The latter are built from quadratic estimators (QEs), which rely on correlations between unequal modes of the lensed CMB anisotropies. This type of estimator was first used for delensing real data in~\cite{carron_internal_2017}, and for constraining cosmological parameters with delensed CMB power spectra in~\cite{han_atacama_2020}. QEs are strongly correlated with the true lensing field on large scales, but remain limited by statistical reconstruction noise on intermediate and small scales. Iterative methods for internal lensing reconstruction are expected to outperform QEs for future very deep high-resolution surveys~\cite{belkner_cmb-s4_2023}. However, at the sensitivity levels of ACT and SO, including external LSS tracers (as proposed in~\cite{sherwin_delensing_2015}) can partially compensate for the higher reconstruction noise, leading to significant improvements in delensing power~\cite{namikawa_simons_2022}. 

One such LSS tracer is the cosmic infrared background (CIB), radiation originating from dusty star-forming galaxies with a broad redshift distribution peaking around $z\sim2$. The significant overlap between the CIB and CMB lensing redshift kernels makes the CIB a good proxy for $\kappa$; examples of its use for delensing are found in~\cite{larsen_demonstration_2016},~\cite{manzotti_cmb_2017} and~\cite{planck_collaboration_planck_2020_lensing}. Maps of galaxy overdensities have also been shown to contribute non-negligible delensing power despite currently probing a narrower redshift range~\cite{yu_multitracer_2017}. In this work, we optimize the delensing performance of our template by combining the ACT lensing reconstruction with a CIB map from Planck and two tomographic galaxy samples from the unWISE catalogue\footnote{Luminous Red Galaxy (LRG) samples from the Dark Energy Spectroscopic Instrument (DESI) were also considered in Appendix~\ref{appendix_C}, but did not add significant delensing power}~\cite{schlafly_unwise_2019}.

This paper is organized as follows. In Sec.~\ref{section:theory}, we review the theoretical foundations of CMB lensing and introduce our multi-tracer delensing framework. Section~\ref{section:data} characterizes the datasets included in our analysis, while Sec.~\ref{section:sims} describes our simulation suites. After outlining our template-construction methodology in Sec.~\ref{section:methodology}, we present our results in Sec.~\ref{section:results}. In particular, we analyze the delensing performance of our template on Planck and ACT DR6 $B$-modes. We then conclude in Sec.~\ref{section:conclusion}. Finally, Appendices~\ref{appendix_C},~\ref{appendix_A} and~\ref{appendix_B} discuss, respectively, the inclusion of additional galaxy samples, technicalities related to the treatment of transfer functions, and internal delensing biases.

\section{Theoretical framework}\label{section:theory}

\subsection{Weak lensing of the CMB}

Weak gravitational lensing deflects the paths of CMB photons, introducing characteristic distortions into the observed temperature and polarization anisotropies. A review of this effect is in Ref.~\cite{lewis_weak_2006}; here we summarize the relevant details. 

In the Born approximation, small deviations produced by gravitational (Weyl) potentials $\Psi$ are integrated over comoving distance $\chi$ along the unlensed line of sight $\hat{\bm{n}}$; in a flat universe, the total deflection angle is given by~\cite{lewis_weak_2006}
\begin{equation}\label{eq:lens_potential}
    \bm{\alpha}=-2\int_{0}^{\chi_{*}}{d\chi \frac{\chi_{*}-\chi}{\chi_{*}\chi}\nabla\Psi\left(\chi\hat{\bm{n}};\eta_0-\chi\right)}.
\end{equation}
Here, $\chi_{*}$ corresponds to the comoving distance of the CMB last-scattering surface, $\eta_0-\chi$ is the conformal time at which the photon was at position $\chi\hat{\bm{n}}$, and $\nabla$ represents the covariant derivative on the sphere. Equation~\eqref{eq:lens_potential} can be simply rewritten as a pure gradient $\bm{\alpha}=\nabla\phi$, defining the lensing potential $\phi$.

For scalar fields such as temperature $\Theta$, the lensing-induced remapping is given at leading order by
\begin{equation}\label{eq:theta_remap}
    \tilde{\Theta}\left(\hat{\bm{n}}\right) = \Theta\left(\hat{\bm{n}}+\bm{\alpha}\right) = \Theta\left(\hat{\bm{n}}\right)+\nabla_i\phi\nabla^i\Theta\left(\hat{\bm{n}}\right)+O\left(\phi^2\right),
\end{equation}
where the tilde refers to lensed quantities. An analogous expression can be obtained for linear polarization, which is encoded by a rank-2 traceless symmetric tensor $\mathcal{P}_{ab}$:
\begin{equation}\label{eq:pol_remap}
    \tilde{\mathcal{P}}_{ab}\left(\hat{\bm{n}}\right) = \mathcal{P}_{ab}\left(\hat{\bm{n}}\right)+\nabla_i\phi\nabla^i\mathcal{P}_{ab}\left(\hat{\bm{n}}\right)+O\left(\phi^2\right).
\end{equation}
In this case, the tensor field is parallel-transported along the geodesic defined by the deflection vector $\bm{\alpha}$ at $\bm{\hat{n}}$~\cite{challinor_geometry_2002}.

Considering a right-handed orthonormal basis $\left(\bm{\hat{n}},\bm{\hat{e}}_1,\bm{\hat{e}}_2\right)$, we can extract the spin-2 components of $\mathcal{P}_{ab}$ as follows:
\begin{equation}\label{eq:spin2}
    \mathcal{P}^{ab}=\frac{1}{2}\left[{_{-2}P\left(\hat{\bm{n}}\right)}\bm{e}_+^{a}\bm{e}_+^{b} + {_{+2}P\left(\hat{\bm{n}}\right)}\bm{e}_-^{a}\bm{e}_-^{b}\right],
\end{equation}
with $\bm{e}_{\pm}=2^{-1/2}\left(\bm{\hat{e}}_1\pm i\bm{\hat{e}}_2\right)$. For the specific choice of $\bm{\hat{e}}_1=\bm{\hat{e}}_{\theta}$ and $\bm{\hat{e}}_2=\bm{\hat{e}}_{\phi}$ along the $\theta$ and $\phi$ directions of a spherical coordinate system, these functions define the Stokes parameters $_{\pm2}P=Q\pm iU$~\cite{okamoto_cmb_2003}. 

The spherical harmonic expansion usually applied to scalar fields such as $\Theta$ and $\phi$ can be extended to polarization using the spin-weight formalism developed in~\cite{goldberg_spin_1967} and~\cite{lewis_analysis_2001}; for $_0X =\Theta$ and $_{\pm2}X = {_{\pm2}P}$, we have
\begin{equation}\label{eq:harmonic_transform}
    _{s}X(\hat{\bm{n}})=\sum_{lm}{_{s}X_{lm}}{_{s}Y_{lm}(\hat{\bm{n}})}.
\end{equation}
The lensing-induced shifts $\delta {_sX_{lm}}={_s\tilde{X}_{lm}}-{_sX_{lm}}$ in temperature and polarization are then obtained by taking the spin-weighted harmonic transforms of Eqs~\eqref{eq:theta_remap} and~\eqref{eq:pol_remap}, respectively. Integrating by parts and using the known analytical result for the triple integral of $_{s}Y_{lm}(\hat{\bm{n}})$ leads to
\begin{equation}\label{eq:lensing_shift}
\setlength\arraycolsep{0.9pt}
    \delta {_sX_{lm}}=(-1)^m\sum_{LM\, l'm'}{\left(\begin{matrix} l & l'& L \\ m & -m' & -M \end{matrix}\right){F^{(s)}_{lLl'}}{_{s}X_{l'm'}\phi_{LM}}},
\end{equation}
where $L$ and $M$ label the spherical-harmonic multipole and azimuthal indices of the lensing potential $\phi$, and the mode-coupling function $F^{(s)}_{lLl'}$ is defined in~\cite{okamoto_cmb_2003}.

Under parity, harmonic coefficients of the spin-2 polarization field transform as ${_{\pm2}P_{lm}} \rightarrow (-1)^l{_{\mp2}P_{lm}}$. It is thus straightforward to extract the parity-even ($E$-mode) component $E_{lm}=2^{-1}\left({_{+2}P_{lm}}+{_{-2}P_{lm}}\right)$ and its parity-odd ($B$-mode) counterpart $B_{lm}=(2i)^{-1}\left({_{+2}P_{lm}}-{_{-2}P_{lm}}\right)$. Assuming negligible primordial tensor perturbations, the lensing contribution to $B$-mode anisotropies then directly follows from Eq.~\eqref{eq:lensing_shift}; one obtains
\begin{equation}\label{eq:lensing_B}
\setlength\arraycolsep{0.9pt}
    B^{\textrm{lens}}_{lm}=(-1)^m\sum_{Ll'Mm'}{\left(\begin{matrix} l & l'& L \\ m & -m' & -M \end{matrix}\right){F^{(2)}_{lLl'}}p^{-}E_{l'm'}\phi_{LM}},
\end{equation}
where $p^{-}$ is 0 for $l+l'+L$ even and $-i$ for $l+l'+L$ odd.

The main implication of this result is that the weak lensing of primary CMB $E$-modes generates secondary $B$-modes, increasing cosmic variance and limiting the precision of constraints on PGW. To mitigate this effect, we construct a gradient-order template of the lensing $B$-mode realization in our sky by substituting empirical estimates of the $E$-modes and lensing potential into Eq.~\eqref{eq:lensing_B}. Due to the mode-coupling encoded by $F^{(2)}_{lLl'}$, delensing the large-scale $B$-modes relevant for inflationary physics requires accurate measurements of $E_{lm}$ and $\phi_{LM}$ at intermediate and small scales~\cite{lewis_weak_2006}. Note that while the derivation of Eq.~\eqref{eq:lensing_B} involves unlensed $E$-modes, we do not have access to these in practice. However, using $\hat{E}^{\rm{WF}}$ is not expected to compromise delensing efficiency: in fact, gradient-order templates built from lensed $E$-modes lead to cancellations between higher-order terms in the $B$-mode residuals~\cite{lizancos_limitations_2021}, and generally yield a better delensing performance than non-perturbative methods. 

\subsection{Internal lensing reconstruction}\label{section:internal_rec}

In this section, we explain how the lensing potential can be reconstructed internally from high-resolution CMB maps, following the description in Ref.~\cite{okamoto_cmb_2003}. Let $\tilde{X}$ and $\tilde{Y}$ be two lensed temperature or polarization fields; at first order and for a fixed $\phi$ realization, their harmonic-space covariance is given by
\begin{multline}\label{eq:off_diag}
    \langle \tilde{X}_{lm} \tilde{Y}_{l’m’} \rangle = \delta_{ll'}\delta_{m-m'}(-1)^{m}C_{l}^{XY} \\ + \sum_{LM}(-1)^M{\left(\begin{matrix} l & l'& L \\ m & m' & -M \end{matrix}\right)f^{XY}_{lLl'}\phi_{LM}},
\end{multline}
where the response functions $f^{XY}_{lLl'}$ are defined in~\cite{okamoto_cmb_2003} and the angle brackets represent an ensemble average over unlensed CMB anisotropies. While the first term in Eq.~\eqref{eq:off_diag} simply reflects the unlensed power spectrum, the second term results from the lensing-induced couplings in Eq.~\eqref{eq:lensing_shift} and can be nonzero for $l\neq l'$. 

An estimator of $\phi$ can then be extracted from these off-diagonal correlations. Defining the diagonal\footnote{We adopt diagonal filtering here, as in the ACT DR6 analysis~\cite{qu_2024}. We note that, even for an isotropic survey, this is slightly sub-optimal as it ignores the $\Theta$-$E$ correlation in the filtering~\cite{maniyar_quadratic_2021} However, it has the virtue that temperature and polarization maps are then separately filtered.} inverse-variance filtering $\bar{X}_{lm} = \hat{X}_{lm}/C_l^{\hat{X}\hat{X}}$ (where $C_l^{\hat{X}\hat{X}}$ is the power spectrum of the noisy observed field $\hat{X}$), Ref.~\cite{okamoto_cmb_2003} introduced the form
\begin{equation}\label{eq:quad_estimator}
\setlength\arraycolsep{0.9pt}
    \hat{\phi}^{XY}_{LM} = A_L^{XY}\sum_{ll'mm'}\left(\begin{matrix} l & l'& L \\ m & m' & -M \end{matrix}\right)\frac{(-1)^M(f^{XY}_{lLl'})^{*}}{\Delta^{XY}}\bar{X}_{lm}\bar{Y}_{l'm'},
\end{equation}
known as a quadratic estimator (QE). Here, $\Delta^{XY}\big|_{X\neq Y}=1$, $\Delta^{XY}\big|_{X=Y}=2$, and the normalization
\begin{equation}\label{eq:normalization}
    A_L^{XY} = (2L+1)\left[\sum_{ll'}{\frac{|f^{XY}_{lLl'}|^2}{\Delta^{XY}C_l^{\hat{X}\hat{X}}C_{l'}^{\hat{Y}\hat{Y}}}}\right]^{-1}
\end{equation}
is set so that $\langle\hat{\phi}^{XY}_{LM}\rangle=\phi_{LM}$. The form of Eq.~\eqref{eq:quad_estimator} guarantees that the diagonal term in Eq.~\eqref{eq:off_diag} does not contribute to $\langle\hat{\phi}_{LM}\rangle$ for $L>0$. Evaluating $\langle\hat{\phi}^{XY}_{LM}\hat{\phi}^{XY*}_{L'M'}\rangle=\delta_{LL'}\delta_{MM'}\left(C_L^{\phi\phi}+N_L^{XY}\right)$ for our QE, the power spectrum of the reconstruction noise, $N_L^{XY}$, arises from the disconnected four-point function of the CMB fields (i.e., from chance Gaussian fluctuations that mimic the effects of lensing).
We can then build a minimum-variance (MV) estimator as a linear combination
\begin{equation}\label{eq:mv_qe}
    \hat{\phi}^{\rm{MV}}_{LM}=\sum_{XY}w_L^{XY} \hat{\phi}^{XY}_{LM},
\end{equation}
where $XY\in\{\Theta\Theta,\Theta E,EE,EB\}$ and the weights $w_L^{XY}$ are chosen to minimize the overall reconstruction noise. Note that an alternative, more optimal approach to constructing a global minimum-variance QE was proposed in~\cite{maniyar_quadratic_2021}.

To reduce computational costs associated with the harmonic-space convolution in Eq.~\eqref{eq:quad_estimator}, the ACT DR6 lensing map (described in detail in Sec.~\ref{section:ACT}) is based on an approximately equivalent real-space QE formulation. This approach consists of estimating the spin-1 components of the deflection vector
\begin{equation}\label{eq:alpha_spin1}
    \bm{\alpha} =\frac{1}{\sqrt{2}}\left[{_{-1}\alpha}\left(\hat{\bm{n}}\right)\bm{e}_+ + {_{+1}\alpha}\left(\hat{\bm{n}}\right)\bm{e}_-\right],
\end{equation}
with harmonic coefficients ${_{\pm1}\alpha_{LM}=\mp\sqrt{L(L+1)}\phi_{LM}}$. 

In~\cite{qu_2024}, an unnormalized displacement-like spin-1 estimator is obtained from the following combination of real-space maps ${_s}X\in\{\Theta,{_{+2}P},{_{-2}P}\}$:
\begin{equation}\label{eq:realspace_QE}
    {_{+1}\hat{d}\left(\hat{\bm{n}}\right)}=-\sum_{s=0,\pm2}{_{-s}\bar{X}\left(\hat{\bm{n}}\right)}\left[\eth {_sX^{\rm{WF}}}\right]\left(\hat{\bm{n}}\right).
\end{equation}
Here, $\eth$ represents the spin-raising operator; its action on the Wiener-filtered fields is explicitly given by
\begin{equation}\label{eq:gradients}
    \left[\eth {_sX^{\rm{WF}}}\right]\left(\hat{\bm{n}}\right)=\sum_{lm}\sqrt{(l-s)(l+s+1)}{}_s X^{\rm{WF}}_{lm}{_{s+1}Y_{lm}\left(\hat{\bm{n}}\right)}
\end{equation}
with ${_{\pm2}X^{\rm{WF}}_{lm}=E^{\rm{WF}}_{lm}\pm iB^{\rm{WF}}_{lm}}$. The Wiener-filtering operation is performed using a set of lensed CMB power spectra $C_l^{XY}$ generated in our fiducial cosmological model, and is expressed in harmonic space as
\begin{equation}\label{eq:wiener_matrix}
    \left(\begin{matrix} \Theta_{lm}^{\rm{WF}} \\ E_{lm}^{\rm{WF}} \\ B_{lm}^{\rm{WF}} \end{matrix}\right) = \left(\begin{matrix} C_l^{\Theta\Theta} & C_l^{\Theta E} & 0 \\ C_l^{\Theta E} & C_l^{E E} & 0 \\ 0 & 0 & C_l^{BB} \end{matrix}\right)\left(\begin{matrix} \bar{\Theta}_{lm} \\ \bar{E}_{lm} \\ \bar{B}_{lm} \end{matrix}\right).
\end{equation}

Decomposing the displacement vector into its gradient and curl components
\begin{equation}\label{eq:grad_curl}
    {_{+1}\hat{d}\left(\hat{\bm{n}}\right)}=-\sum_{LM}\frac{\hat{g}_{LM}+i\hat{c}_{LM}}{\sqrt{L(L+1)}}{_{1}Y_{LM}\left(\hat{\bm{n}}\right)},
\end{equation}
we can finally extract a proxy for the lensing potential
\begin{equation}\label{eq:phi_norm}
    \hat{\phi}_{LM}=\mathcal{R}_L^{-1}\hat{g}_{LM}.
\end{equation}
Again, the normalization factor $\mathcal{R}_L^{-1}$ is fixed so that $\langle\hat{\phi}_{LM}\rangle=\phi_{LM}$; as a result of the Wiener filtering in Eq.~\eqref{eq:realspace_QE}, this estimator is already minimum-variance and is approximately equivalent to $\hat{\phi}^{\textrm{MV}}_{LM}$ defined in Eq.~\eqref{eq:mv_qe}~\cite{qu_2024}. Note that the curl component of Eq.~\eqref{eq:grad_curl} is expected to be zero and can be used as a null test.

To build a temperature-only estimator, we select the $s = 0$ part of Eq.~\eqref{eq:grad_curl} and set $\bar{E}_{lm}=0$ in Eq.~\eqref{eq:wiener_matrix}. Conversely, the polarization-only variant uses the $s=\pm2$ terms with $\bar{\Theta}_{lm}=0$ in the Wiener filter. Finally, all ACT lensing reconstructions are converted into maps of the convergence field $\kappa=-\nabla^2\phi/2$.

\subsection{Large-scale structure tracers}\label{subsec:lss_theory}

At the ACT sensitivity level, the QE reconstruction noise described by Eq.~\eqref{eq:normalization} (sourced by statistical fluctuations of the unlensed CMB) becomes dominant over $C_L^{\phi\phi}$ for $L\gtrsim200$~\cite{qu_2024}. As the intermediate and small scales $200 \leq L \leq 800$ are particularly relevant for delensing large-scale $B$-modes, this motivates the complementary use of external LSS tracers. The optimal linear combination of tracers is determined by characterizing the correlation of each individual dataset with the true lensing field and estimating the resulting delensing power.

Similarly to the CMB lensing convergence, the maps of the CIB and of the galaxy overdensity we use here are 2D fields on the sphere which probe the 3D distribution of matter in the Universe. Defining the dimensionless matter overdensity $\delta_m=\left(\rho_m-\bar{\rho}_m\right)/\bar{\rho}_m$ (the bar denotes average quantities), such projected fields can be expressed as a line-of-sight integral
\begin{equation}\label{eq:field_projection}
    X\left(\hat{\bm{n}}\right) = \int dz W_X(z) {\delta_m\left(\chi(z)\hat{\bm{n}},z\right)},
\end{equation}
where $\chi(z)$ represents the comoving distance back to redshift $z$ and the kernel $W_X(z)$ encodes the relative contribution of sources at $z$.

\begin{figure}[h]
    \centering
    \includegraphics[width=\linewidth, height=0.25\textheight]{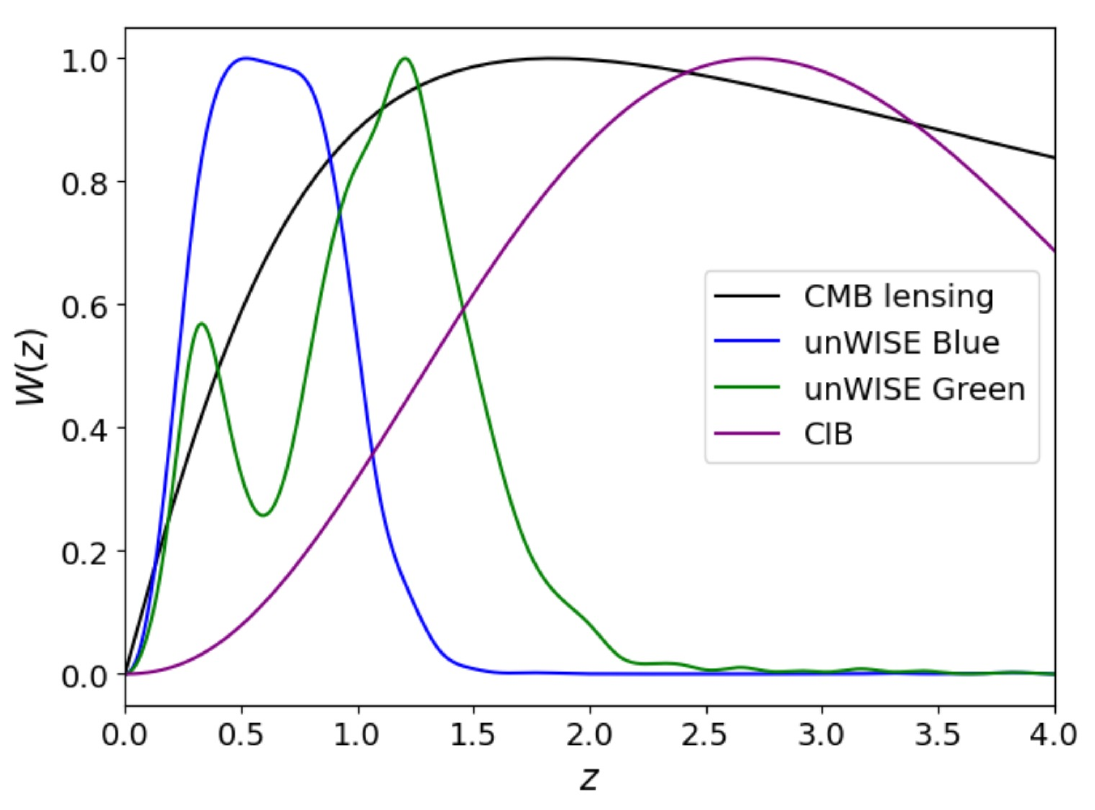}
    \vspace{-15pt}
    \caption{Normalized projection kernel as a function of redshift for each LSS tracer used in our analysis: the CMB lensing convergence (Eq.~\ref{eq:lens_kernel}; black), the unWISE Blue and Green galaxy samples (Eq.~\ref{eq:galaxy_kernel}, Sec.~\ref{section:unWISE}, Ref.~\cite{schlafly_unwise_2019}; blue and green lines, respectively) and the CIB (Eq.~\ref{eq:cib_kernel}, Sec.~\ref{section:CIB}; purple).}
    \label{fig:kernels}
\end{figure}

The CMB lensing kernel is computed using \mbox{$\kappa=-\bm{\nabla}\cdot\bm{\alpha}/2$} and substituting Eq.~\eqref{eq:lens_potential}. The resulting integral involves the angular Laplacian $\nabla^2\Psi$ of the gravitational potential, which is linked to the matter overdensity by the cosmological Poisson equation 
\begin{equation}\label{eq:poisson}
    \nabla^2\Psi\approx \chi^2\nabla^2_{\chi}\Psi=4\pi G \chi^2(1+z)^{-2}\bar{\rho}_m\delta_m.
\end{equation}
Here, $\nabla^2_{\chi}$ is the full Laplacian at comoving distance $\chi$, and the background matter density is given by \mbox{$\bar{\rho}_m=3H_0^2\Omega_m(1+z)^3/8\pi G$} with $H_0$ the Hubble constant. We have dropped the radial derivatives, which is an excellent approximation given the broad redshift kernel (see, e.g.,~\cite{gao_2024}). Noting the relation $d\chi=dz/H(z)$, where $H(z)$ is the Hubble parameter and we set $c=1$, one obtains
\begin{equation}\label{eq:lens_kernel}
    W_{\kappa}(z)=\frac{3}{2}\Omega_m H_0^2\chi(z)\frac{(1+z)}{H(z)}\frac{\chi_{*}-\chi(z)}{\chi_{*}}
\end{equation}
in a spatially flat Universe. This kernel is shown as the black line in Fig.~\ref{fig:kernels}; it peaks at $z\sim 2$ and falls off slowly, extending to high redshifts.

Galaxy surveys cover a more limited redshift range, with the unWISE samples used in this work spanning $0.2 \leq z \leq 1.8$ (see Sec.~\ref{section:unWISE} for a detailed description). The galaxy density field is a biased tracer of the underlying dark matter distribution: galaxies tend to form in overdense regions, leading to an enhancement in the clustering signal. On the scales of interest when delensing large-scale $B$-modes, this effect is approximated by a linear bias factor $b(z)$, which we parametrize here as $b(z)=b_0+b_1z$. Note that galaxy bias is scale-dependent in reality; however, this is expected to have limited impact on delensing~\cite{ref:wang_et_al_25}. The projection kernel 
\begin{equation}\label{eq:galaxy_kernel}
    W_g(z)=b(z)\frac{dN/dz}{\int dz'dN/dz'}
\end{equation}
then directly follows from the normalized redshift distributions, which are plotted for our unWISE maps as the blue and green lines in Fig.~\ref{fig:kernels}.

For the CIB, we consider the (comoving) emissivity density $j(\nu,z)$ as a biased tracer of dark matter, with $\delta_j=\left(j-\bar{j}\right)/\bar{j}= b_c\delta_m$. The single spectral energy distribution (SED) model in~\cite{hall_angular_2010} describes the dust emission spectrum as a modified blackbody transitioning into a power-law decline at a rest-frame frequency $\nu'\sim4955$~GHz. Mathematically, the SED is given by
\begin{equation}\label{eq:dust_sed}
f_{\nu}\propto
    \begin{cases}
     \nu^{\beta}B_{\nu}\left(T_d\right) & \text{for }\nu\le \nu'\\    
    \left(\nu/\nu'\right)^{-\alpha} & \text{for }\nu > \nu',   
    \end{cases}
\end{equation}
where $B_{\nu}\left(T_d\right)$ is the Planck function; $f_{\nu}$ is assumed to be identical for all source galaxies. The effective dust temperature $T_d=34\,\text{K}$ and spectral indices $\beta=\alpha=2$ are chosen in agreement with observations~\cite{hall_angular_2010}. Considering a Gaussian distribution of emissivity around $z_c\sim2$, the mean comoving emissivity density then corresponds to
\begin{equation}\label{eq:cib_jnu}
    \bar{j}(\nu,z)\propto \chi^2e^{-\frac{(z-z_c)^2}{2\sigma_z^2}}(1+z)^{-1}f_{\nu(1+z)},
\end{equation}
where we follow~\cite{yu_multitracer_2017} and set $\sigma_z=2$. With this definition, the CIB intensity variations are given by $\delta I_{\nu}=\int dz\,H(z)^{-1}(1+z)^{-1}\bar{j}(\nu,z)\delta_j$~\cite{hall_angular_2010}, leading to the projection kernel~\cite{yu_multitracer_2017}
\begin{equation}\label{eq:cib_kernel}
    W_I(z)=b_c\frac{\chi^2(z)}{H(z)(1+z)^2}e^{-\frac{(z-z_c)^2}{2\sigma_z^2}}f_{\nu(1+z)}.
\end{equation}
This kernel, shown in purple in Fig.~\ref{fig:kernels}, is complementary to that of unWISE as it has significant overlap with $W_{\kappa}(z)$ at redshifts above the range typically probed by galaxy surveys.

To quantify the delensing power of a specific tracer $\hat{\kappa}^i$, we evaluate its correlation coefficient
\begin{equation}\label{eq:corr_coeff}
    \rho^{\kappa \hat{\kappa}^i}_L=\frac{C_L^{\kappa \hat{\kappa}^i}}{\sqrt{C_L^{\kappa\kappa}C_L^{\hat{\kappa}^i\hat{\kappa}^i}}}
\end{equation}
with the true lensing convergence. For $L\gtrsim100$, fluctuations transverse to the line of sight dominate and the power spectra in Eq.~\eqref{eq:corr_coeff} can be computed in the Limber approximation~\cite{limber_analysis_1953}. Within this framework, the cross-spectrum of two projected fields $X\left(\hat{\bm{n}}\right)$ and $Y\left(\hat{\bm{n}}\right)$ defined as in Eq.~\eqref{eq:field_projection} is simplified as follows:
\begin{equation}\label{eq:limber}
    C_L^{X Y}
    \approx\int dz\, H(z) \frac{W_X(z)W_Y(z)}{\chi^2(z)}P\left(k=\frac{L+1/2}{\chi(z)},z\right),
\end{equation}
where $P(k,z)$ is the (dimensional) matter power spectrum.

The main takeaway of Eqs~\eqref{eq:corr_coeff} and~\eqref{eq:limber} is that the correlation between a given tracer and the true $\kappa$ increases the more the tracer overlaps in redshift and scale with the relevant matter modes, and is limited by the noise level in the auto-spectrum $C_L^{\hat{\kappa}^i\hat{\kappa}^i}$. Combining multiple datasets with complementary projection kernels and different noise properties therefore allows us to surpass the delensing performance of individual tracers. In practice, the optimal estimator $\hat{\kappa}^{\textrm{comb}}$ is a Wiener filter, constructed as a weighted sum of all considered tracers with coefficients determined in harmonic space to maximize $\rho_L^{\kappa\hat{\kappa}^{\textrm{comb}}}$. Ignoring correlations between multipoles\footnote{Further optimization to account for such effects was done in Ref.~\cite{namikawa_litebird_2023}. The simpler expression used in the present work, which is optimal for isotropic surveys, is expected to be close to optimal for the relatively large sky fraction covered by ACT.}, the solution to this optimization problem is given by~\cite{sherwin_delensing_2015,yu_multitracer_2017}
\begin{equation}\label{eq:kappa_comb}
    \hat{\kappa}^{\textrm{comb}}_{LM}=\sum_{ij}(\bm{\rho_{L}}^{-1})^{ij}\rho^{\hat{\kappa}^j\kappa}_{L}\sqrt{\frac{C_L^{\kappa\kappa}}{C_L^{\hat{\kappa}^{i}\hat{\kappa}^{i}}}}\hat{\kappa}^{i}_{LM},
\end{equation}
where we define the matrix $(\bm{\rho_{L}})^{ij}=\rho_L^{\hat{\kappa}^{i}\hat{\kappa}^{j}}$.

\subsection{Delensing performance}

The final step of our analysis consists of building a model of the lensing $B$-modes based on observed $E$-modes and on the convergence estimator in Eq.~\eqref{eq:kappa_comb}; this section investigates the expected delensing performance of such a template.

Noise in the $E$-mode data is mitigated by applying a diagonal Wiener filter
\begin{equation}\label{eq:wiener}
\hat{E}_{lm}^{\rm{WF}} = \frac{C_l^{EE}}{C_l^{EE}+N_l^{EE}} \hat{E}_{lm},
\end{equation}
where hats refer to observed quantities and the noise power spectrum $N_l^{EE}$ is estimated from splits of the ACT maps (see Sec.~\ref{section:ACT}). Note that this expression is equivalent to Eq.~\eqref{eq:wiener_matrix} with $\bar{\Theta}_{lm}=0$.\footnote{A more complex filtering operation accounting for masking and noise inhomogeneities is introduced in~\cite{eriksen_power_2004} and performed in~\cite{namikawa_simons_2022}. As the ACT noise levels are relatively uniform across the survey area, we use the simpler diagonal filter to reduce computational costs.}

We obtain our lensing $B$-mode template by evaluating Eq.~\eqref{eq:lensing_B} for $\hat{E}^{\rm{WF}}$ and $\hat{\kappa}^{\textrm{comb}}$ before converting the resulting harmonic coefficients into real-space $Q$ and $U$ maps. Delensing can then be performed at the map level by directly subtracting the template from the observed $B$-modes. Though conceptually simple, this approach is more sensitive to survey non-idealities such as masking, filtering, and beam systematics. An alternative method, demonstrated in~\cite{namikawa_simons_2022, the_simons_observatory_collaboration_simons_2024}, extends the usual power-spectrum-based likelihood for $r$ inference to include all cross-spectra between the template and the observed multi-frequency CMB maps, effectively treating the template as an additional data channel. These two approaches were shown to yield equivalent constraints on $r$, both in idealized analytical forecasts and in a full mock analysis using SO-like simulations~\cite{the_simons_observatory_collaboration_simons_2024}.

For a statistically isotropic survey, a first-order approximation of the ensemble-averaged auto- and cross-spectra of the lensing $B$-mode template can be derived using the following property:
\begin{equation}\label{eq:kappa_comb_spectra}
\langle C_L^{\kappa\hat{\kappa}^{\textrm{comb}}} \rangle \!=\!\langle C_L^{\hat{\kappa}^{\textrm{comb}}\hat{\kappa}^{\textrm{comb}}}\rangle\!=\!C_L^{\kappa\kappa}\sum_{ij}\rho^{i\kappa}_{L}(\rho^{-1})^{ij}_{L}\rho^{j\kappa}_{L}.
\end{equation}
This relation is a consequence of the weights chosen in Eq.~\eqref{eq:kappa_comb}. Convolving with the Wiener-filtered $E$-modes and assuming the noise to be uncorrelated with the CMB signal ($C_l^{E\hat{E}}=C_l^{EE}$), the cross-spectrum between the template $B_t$ and the true lensing $B$-modes is given by
\begin{align}
    C^{B_{t}B}_l &= \sum_{l'L}{|\mathcal{M}(l,l',L)|^2 \frac{C^{EE}_{l'}}{C^{\hat{E}\hat{E}}_{l'}}C^{E\hat{E}}_{l'}
    C^{\kappa\hat{\kappa}^{\textrm{comb}}}_{L}} \nonumber \\
    &= \sum_{l'L}{|\mathcal{M}(l,l',L)|^2 \frac{C^{EE}_{l'}}{C^{\hat{E}\hat{E}}_{l'}}C^{EE}_{l'}C^{\kappa\hat{\kappa}^{\textrm{comb}}}_{L}}, \label{eq:Cl_Bxtemp}
\end{align}
where $\mathcal{M}(l,l',L)$ encodes the prefactors from Eq.~\eqref{eq:lensing_B}. As a result of the optimal tracer weighting and $E$-mode filtering, Eq.~\eqref{eq:Cl_Bxtemp} is identical to the lensing template auto-spectrum 
\begin{align}
    C^{B_{t}B_{t}}_l &= \sum_{l'L}{|\mathcal{M}(l,l',L)|^2 \left(\frac{C^{EE}_{l'}}{C^{\hat{E}\hat{E}}_{l'}}\right)^2C^{\hat{E}\hat{E}}_{l'}C^{\hat{\kappa}^{\textrm{comb}}\hat{\kappa}^{\textrm{comb}}}_{L}} \nonumber \\
    &= \sum_{l'L}{|\mathcal{M}(l,l',L)|^2 \frac{C^{EE}_{l'}}{C^{\hat{E}\hat{E}}_{l'}}C^{EE}_{l'}C^{\kappa\hat{\kappa}^{\textrm{comb}}}_{L}}. \label{eq:Cl_tempxtemp}
\end{align}

Expressing the delensed $B$-modes as $B^{\rm{del}}_{lm}=\tilde{B}_{lm}-B^t_{lm}$, we use the equivalence of Eqs~\eqref{eq:Cl_Bxtemp} and~\eqref{eq:Cl_tempxtemp} to estimate the remaining power after delensing:
\begin{equation}\label{eq:delensed_B}
    C_l^{BB,\rm{del}}\approx C_l^{\tilde{B}\tilde{B}}-C_l^{B_tB_t}.
\end{equation}
While the delensing efficiency $C_l^{B_tB_t}/C_l^{\tilde{B}\tilde{B}}$
is a scale-dependent quantity, its variations with $l$ are generally rather small. We can then define the multipole-averaged fractional residual $B$-mode power
\begin{equation}\label{eq:Alens}
    A_{\rm{lens}} = \overline{1-\frac{C_l^{B_tB_t}}{C_l^{\tilde{B}\tilde{B}}}},
\end{equation}
with $C_l^{BB,\rm{del}}\approx A_{\rm{lens}}C_l^{BB,\textrm{lens}}$.

\section{Data}\label{section:data}

\subsection{ACT DR6 maps}\label{section:ACT}

As internal tracers, we use the publicly available ACT DR6 CMB lensing convergence maps\footnote{The maps and associated products can be downloaded at \url{https://lambda.gsfc.nasa.gov/product/act/actadv_dr6_lensing_maps_get.html}.}. The minimum-variance $\hat{\kappa}$ reconstruction is built from night-time temperature and polarization data collected between 2017 and 2022~\cite{qu_2024}. A $\Theta$-only variant is also available, which we use to avoid mode overlap when delensing ACT $B$-modes in Sec.~\ref{section:results}. The construction of these lensing maps, which we summarize in this section, is presented in full detail in Refs~\cite{qu_2024} and~\cite{madhavacheril_atacama_2024}.

Based in the Atacama Desert in Chile, ACT produced arcminute-resolution CMB maps over around $45\%$ of the sky in five frequency bands. Two of them are used in our analysis: f090 (77–-112\,GHz) and f150 (124-–172\,GHz). Data in each band was divided into four splits with independent noise realizations and mapped separately. Individual frequency maps were then beam-deconvolved and co-added in harmonic space with inverse-variance weighting. Finally, a 2D Fourier-space filter was applied to the maps in order to mitigate ground pickup, removing modes with $\big|l_x \big| < 90$ and $\big |l_y \big| < 50$~\cite{choi_atacama_2020}.

The lensing convergence was extracted from CMB data at $600 < l < 3000$ using the \texttt{so-lenspipe} code developed for SO, which computes the QEs introduced in Sec.~\ref{section:internal_rec}. The lower bound of this multipole range was chosen to mitigate atmospheric noise and Galactic foregrounds; it will also serve to avoid mode overlap between the $EB$ QE and the large-scale $B$-modes we will seek to delens in future studies. The upper bound aimed to limit contamination by extragalactic foregrounds~\cite{lizancos_impact_2022}.

To avoid having to model noise accurately to remove power spectrum biases, the pipeline implemented the split-based estimator $\hat{\phi}_{LM}=\frac{1}{6}\sum_{i<j}\hat{\phi}_{LM}^{(ij)}$ defined in~\cite{madhavacheril_cmb_2021}, where Eq.~\eqref{eq:phi_norm} is evaluated and averaged over pairs of independent data splits $i\neq j$. An additional bias-hardening correction was applied to Eq.~\eqref{eq:phi_norm} to suppress biases related to extragalactic foregrounds such as the CIB and thermal Sunyaev–Zel’dovich (tSZ) clusters. This process, known as profile hardening, is described in more detail in~\cite{maccrann_atacama_2023, madhavacheril_atacama_2024}.

\begin{figure}[h]
    \centering
    \includegraphics[width=\linewidth, height=0.22\textheight]{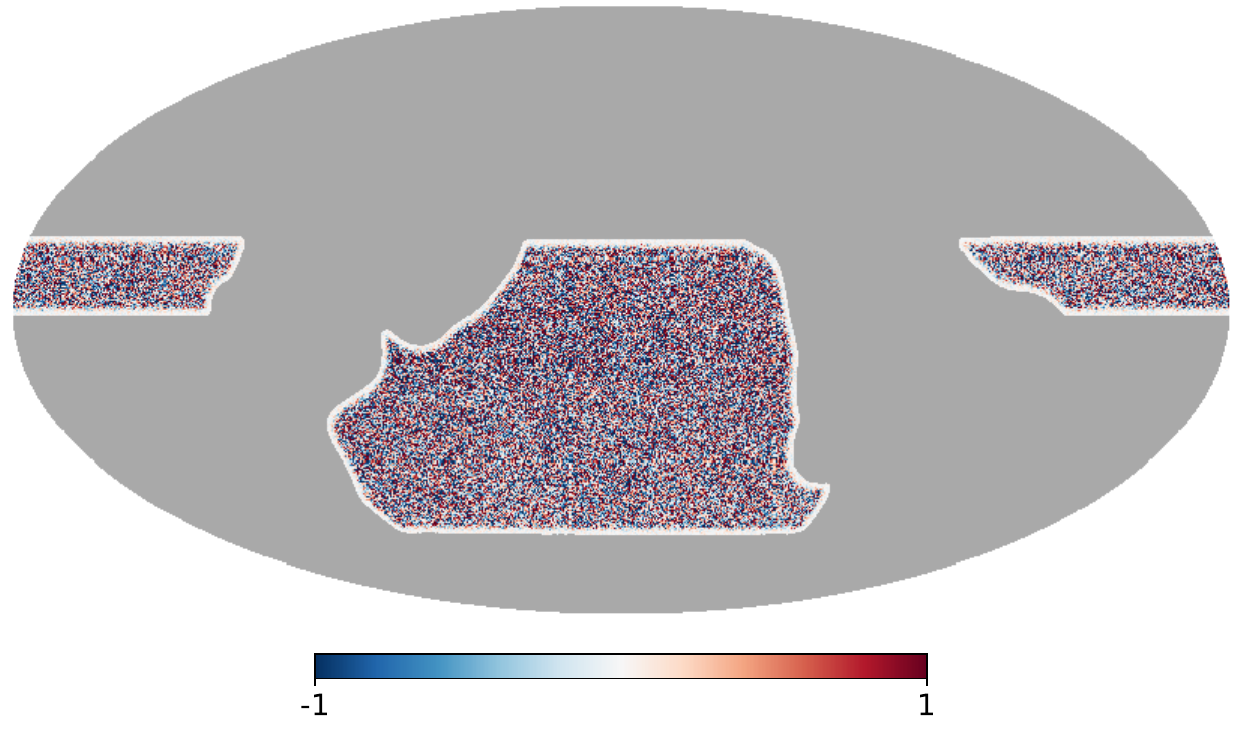}
    \vspace{-5pt}
    \caption{ACT DR6 lensing convergence map (dimensionless) in Equatorial coordinates. This map was produced with the MV quadratic estimator described in Sec.~\ref{section:internal_rec}, using cross-correlations between data splits to mitigate noise biases~\cite{madhavacheril_atacama_2024}. Masked pixels are shown in grey, leaving an unmasked sky fraction $f_{\rm{sky}}= 0.23$.}
    \label{fig:act_lensing}
\end{figure}

Masking was applied to regions of bright Galactic emission based on Planck data at 353\,GHz, as well as to pixels with noise levels above $70\,\mu\text{K-arcmin}$. The mask was then apodized with a 3~deg cosine-squared edge roll-off~\cite{qu_2024}. In addition, bright point sources were masked and inpainted, and detected tSZ clusters were subtracted. The remaining sky fraction $f_{\rm{sky}}= 0.23$, corresponding to an area of about $9400\,\text{deg}^2$, is shown in Fig.~\ref{fig:act_lensing}.

Finally, note that QEs are sensitive to the statistical anisotropy introduced by this partial sky coverage; this effect manifests itself as a nonzero ensemble average $\langle\hat{\phi}_{LM}\rangle$ over realizations of the CMB and lensing potential. This spurious term, known as ``mean-field bias'', was estimated from simulations and subtracted as indicated in Ref.~\cite{qu_2024}.

\subsection{unWISE galaxies}\label{section:unWISE}

The galaxy samples used in this work are extracted from the unWISE  catalogue~\cite{schlafly_unwise_2019,krolewski_unwise_2020}, including data collected by the Wide-Field Infrared Survey Explorer (WISE) satellite during its original one-year mission~\cite{wright_wide-field_2010} and four years of its post-hibernation NEOWISE phase~\cite{mainzer_initial_2014}. With approximately two billion sources observed at 3.4 and $4.6\,\mu\text{m}$ (W1 and W2 bands), this dataset constitutes the deepest full-sky infrared catalogue to date.

We construct galaxy overdensity maps based on the unWISE Blue and Green samples, as defined by the W1-W2 colour cuts listed in Ref.~\cite{schlafly_unwise_2019} and~\cite{krolewski_unwise_2020}. The corresponding normalized redshift distributions, determined by cross-matching with photometric redshifts from the COSMOS2015 catalogue~\cite{laigle_cosmos2015_2016}, are shown in Fig.~\ref{fig:kernels}. The Blue and Green samples have mean redshifts of $0.6$ and $1.1$, respectively, and span the overall range $0.2 \leq z \leq 1.8$. Note that~\cite{schlafly_unwise_2019} also defines a Red sample, which we do not use here due to its significantly lower number density. 

\begin{figure}[h]
    \centering
    \includegraphics[width=\linewidth, height=0.22\textheight]{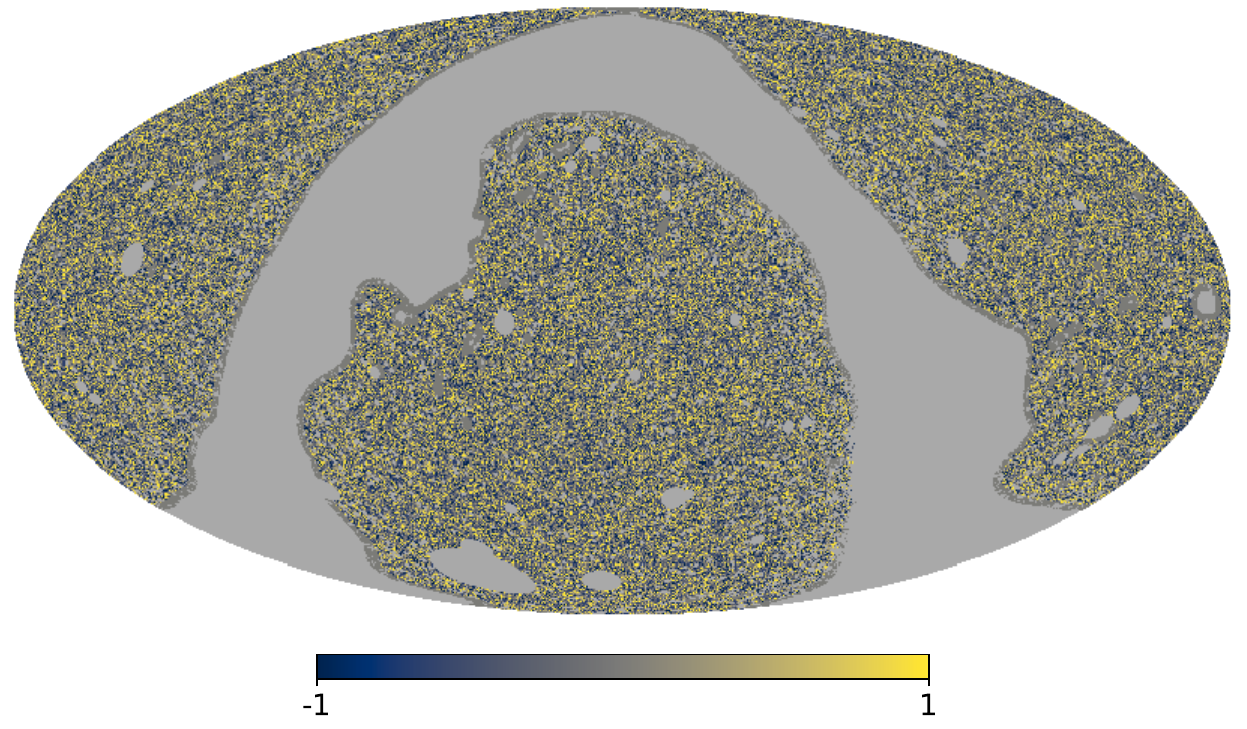}
    \vspace{-5pt}
    \caption{Dimensionless overdensity field as defined in Eq.~\eqref{eq:unWISE_density} for the unWISE Blue galaxy sample. Masked pixels are represented in grey. A similar result, not shown here, is obtained for the Green sample. Both maps are generated in Equatorial coordinates, using the HEALPix equal-area pixelization~\cite{gorski_healpix_2005} with $n_{\rm{side}}=2048$.}
    \label{fig:unwise_blue}
\end{figure}

The large-scale component of the unWISE mask is based on the Planck 2018 lensing mask~\cite{planck_collaboration_planck_2020_lensing} and apodized with a $1\,\text{deg.}$ Gaussian kernel. Bright stars, diffraction spikes and regions within $2.75\,\text{arcsec}$ of a Gaia point source are also removed, yielding a remaining sky fraction $f_{\rm{sky}}=0.59$ (see Fig.~\ref{fig:unwise_blue}). As explained in~\cite{farren_atacama_2023}, these additional cuts result in smaller and variable effective areas for each pixel of the map; measured galaxy number counts are corrected for this effect, and pixels with less than 80\% coverage are masked. 

Finally, we mitigate the two most important imaging systematics by applying weights to remove correlations between galaxy number counts and stellar density as well as survey depth. The construction of these weights is described in full detail in~\cite{farren_atacama_2023}. The galaxy overdensity field is then defined as
\begin{equation}\label{eq:unWISE_density}
    \delta g_i = \frac{g^w_i-\bar{g}^w}{\bar{g}^w},
\end{equation}
where $g^w_i$ represents the weighted (unbiased) number count in pixel $i$ and $\bar{g}^w$ is the average of $g^w_i$ over unmasked pixels. As an example, the Blue sample overdensity map is displayed in Fig.~\ref{fig:unwise_blue}.

\subsection{Planck CIB}\label{section:CIB}

Our final external LSS tracer is a CIB emission map extracted from Planck PR2 data at 353\,GHz with the Generalized Needlet Internal Linear Combination (GNILC) algorithm~\cite{planck_collaboration_planck_2016_1}. This method uses a spherical form of wavelets (known as needlets) to decompose sky maps into different spatial scales, then computes a weighted sum of multifrequency data at each scale to capture the CIB signal while minimizing Galactic foreground residuals and noise contributions~\cite{remazeilles_foreground_2011}.

\begin{figure}[h]
    \centering
    \includegraphics[width=\linewidth, height=0.22\textheight]{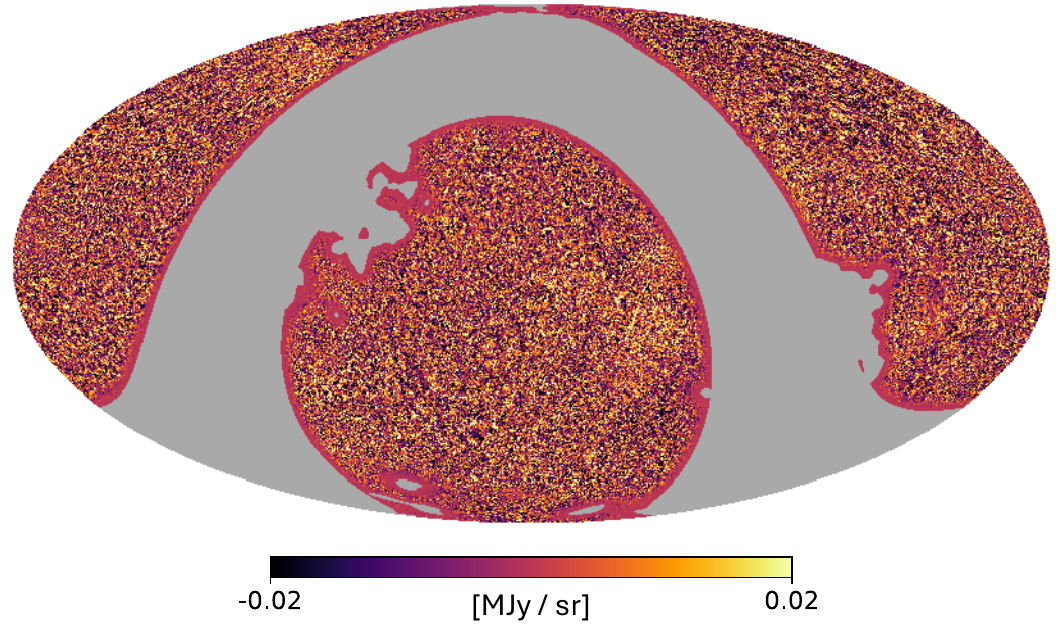}
    \vspace{-5pt}
    \caption{CIB intensity map extracted from Planck 353\,GHz data with the GNILC algorithm. The map is generated at $n_{\rm{side}}=2048$ in Equatorial coordinates, with masked pixels shown in grey.}
    \label{fig:cib_map}
\end{figure}

The resulting CIB map has a resolution equivalent to a 5\,arcmin Gaussian beam, which we deconvolve prior to our analysis. We apply a mask based on the Planck PR2 60\% Galactic cuts with $2\,\text{deg.}$  Gaussian apodization, and further exclude point sources detected at a signal-to-noise ratio greater than five. The unmasked area, covering a sky fraction $f_{\rm{sky}}=0.58$, is displayed in Fig.~\ref{fig:cib_map}.

While GNILC CIB maps are also available at 545 and 857 GHz~\cite{planck_collaboration_planck_2016_1}, we restrict our analysis to the 353\,GHz channel, which has the highest correlation coefficient with the ACT lensing reconstruction. As the other two frequency bands are strongly correlated with the 353\,GHz map, including them would not significantly improve the quality of our combined lensing tracer~\cite{sherwin_delensing_2015}. 

Similarly, the maps produced in ~\cite{mccarthy_large-scale_2024} with the \texttt{pyilc} software were found to be less correlated with CMB lensing than the ones obtained with GNILC in the multipole range of interest. Indeed, despite having the advantage of preserving all CIB power on large scales, this alternative needlet ILC method results in increased dust contamination at intermediate and small scales.

\section{Simulations}\label{section:sims}

For the purpose of estimating statistical errors and understanding the impact of survey non-idealities, we produce 100 mock lensing $B$-mode templates. Their construction requires simulated maps of both the CMB and appropriately correlated external lensing tracers.

\subsection{ACT polarization and lensing}\label{subsec:cmb_sims}

Unlensed CMB temperature and polarization anisotropies are generated on the full sky by drawing Gaussian harmonic coefficients from CAMB-produced~\cite{lewis_efficient_2000} power spectra. Our fiducial cosmology uses the Planck best-fit parameters listed in~\cite{calabrese_cosmological_2017} and assumes $r=0$ (no primordial $B$-modes). Random realizations of a Gaussian lensing potential are obtained in a similar manner, and the \texttt{pixell} package~\cite{naess_pixell_2021} is used to lens the CMB maps.

The complex spatially-varying properties of the atmospheric and instrumental noise affecting ACT observations are directly estimated from the data using differences between independent splits. As described in~\cite{atkins_atacama_2023}, these are accounted for when producing realistic ACT-like noise maps\footnote{The public code used to generate these maps can be found here: \url{https://github.com/simonsobs/mnms}.}. Gaussian foreground realizations, including extragalactic sources and Galactic dust emission, are drawn from power spectra based on the Sehgal~\cite{sehgal_simulations_2010} and Websky~\cite{stein_websky_2020} simulations. Both noise and foregrounds are added to the single-frequency lensed CMB maps, which are then filtered in Fourier space (see Sec.~\ref{section:ACT}) and smoothed with the relevant Gaussian beams. 

The final versions of the simulated sky maps are obtained by performing an inverse-variance co-addition of the beam-deconvolved f90 and f150 bands. These maps, saved at a resolution $n_{\rm{side}}=2048$, are used as inputs for the internal lensing estimation pipeline to generate realistic ACT-like $\kappa$ reconstructions. 

Note that we also produce a set of 100 noise- and foreground-free CMB simulations including Fourier-space filtering and masking; these are used to estimate transfer functions in Sec.~\ref{section:methodology} and Appendix~\ref{appendix_A}.

\subsection{External tracers}

To mimic galaxy overdensity and CIB maps, we simulate Gaussian fields whose correlations with the lensing convergence realizations from Sec.~\ref{subsec:cmb_sims} match the cross-spectra measured between ACT lensing and our external LSS tracers. This is done by applying the algorithm mentioned in~\cite{farren_atacama_2023} and explained in full detail in Appendix F of~\cite{lizancos_delensing_2022}.

The method works as follows: considering a noise-free simulated lensing field $\kappa_{LM}$ (one of the inputs used to lens the CMB simulations), we seek to draw the harmonic coefficients of a tracer $g_{LM}$ such that \mbox{$\langle \kappa_{LM}g^\ast_{L'M'}\rangle=\delta_{LL'}\delta_{MM'}C_L^{g\kappa}$} and \mbox{$\langle g_{LM}^{*}g_{L'M'}\rangle=\delta_{LL'}\delta_{MM'}C_L^{gg}$}. Here, $C_L^{g\kappa}$ and $C_L^{gg}$ are smooth functions fitted to the measured auto- and cross-spectra as explained in Sec.~\ref{section:methodology}. The form
\begin{equation}
    g_{LM}=A_L^{g\kappa}\kappa_{LM}+u_{LM},
\end{equation}
where $u_{LM}$ is a Gaussian field uncorrelated with $\kappa_{LM}$, fulfills both conditions if
\begin{equation}
    A_L^{g\kappa}=\frac{C_L^{g\kappa}}{C_L^{\kappa\kappa}}
\end{equation}
and
\begin{equation}
    C_L^{uu}=C_L^{gg}-\left(A_L^{g\kappa}\right)^2C_L^{\kappa\kappa}.
\end{equation}

The algorithm can be extended to ensure appropriate cross-correlations for an arbitrary number of simulated tracers $\hat{\kappa}^i$ by setting $\hat{\kappa}^{0}_{LM}=u_{LM}^0=\kappa_{LM}$ and performing a Cholesky decomposition
\begin{equation}
C_L^{\hat{\kappa}^i\hat{\kappa}^j} = [G_L G_L^T]^{ij}   
\end{equation}
for each $L$, where $G_L$ is lower-triangular. We then generate the coefficients
\begin{equation}
\hat{\kappa}^i_{LM} = \sum_j G_L^{ij} u^j_{LM},  
\end{equation}
where the $u^j_{LM}$ are independent Gaussian fields with unit power spectra, $\langle u^i_{LM} (u^j_{L'M'})^\ast\rangle = \delta^{ij} \delta_{LL'}\delta_{MM'}$.

In the present work, we use the characteristics of the two unWISE samples and of the Planck CIB to generate three simulated Gaussian tracers for each realization of the input $\kappa$. These and the corresponding ACT-like lensing reconstruction are then co-added in harmonic space and convolved with the $E$-modes extracted from the same realization of the ACT polarization simulations (Sec.~\ref{subsec:cmb_sims}), following the process described in the next section.

\section{Methodology}\label{section:methodology}

The first step of our template construction pipeline consists of determining the optimal weights in Eq.~\eqref{eq:kappa_comb} from the cross-spectra between each pair of LSS tracers. The latter are stored as HEALPix maps with $n_{\rm{side}}=2048$ and their respective masks are shown in Sec.~\ref{section:data}. 

In the pseudo-$C_l$ formalism, the cross-spectrum of two masked scalar fields is given by
\begin{equation}\label{eq:pseudo-Cl}
    \hat{C}^{ab,vw}_L=\left(2L+1\right)^{-1}\sum_{M}{a^v_{LM}(b_{LM}^{w}})^\ast,
\end{equation}
where $a^v_{LM}$ and $b^w_{LM}$ are the harmonic coefficients of the maps multiplied by the weights $v(\hat{\bm{n}})$ and $w(\hat{\bm{n}})$ in real space. For statistically isotropic fields, the expectation value of $\hat{C}_L^{ab,vw}$ is related to the true cross-spectrum $C_L^{ab}$ by a mode-coupling matrix $\mathbf{M}$ such that
\begin{equation}\label{eq:namaster}
\langle \hat{C}^{ab,vw}_L \rangle =\sum_{L'}{\mathbf{M}_{LL'}C_{L'}^{ab}}.
\end{equation}
We use the \texttt{NaMaster}~\cite{alonso_unified_2019} implementation of the MASTER algorithm~\cite{hivon_master_2002} to compute $\mathbf{M}_{LL'}$, deconvolve the pixel window function, and recover unbiased estimates $\hat{C}_{L}^{ab}$ of the (cross-)power spectra for each pair of LSS tracers. Note that the mode-coupling matrix only depends on the masks and that it is in general not invertible as a consequence of information loss from partial sky coverage. The algorithm therefore constructs a smaller matrix $\mathbf{M}_{bb'}$ by binning multipoles into bandpowers $b$, with $L\in b$ if $L_{\rm{min}}+b\Delta L\leq L \leq L_{\rm{min}}+(b+1)\Delta L$. We evaluate the cross-spectra of our LSS tracers over a multipole range $2\leq L \leq 2000$ and set the bin width to $\Delta L=100$.

\begin{figure*}[!t]
    \centering
    \begin{minipage}{.49\textwidth}
        \centering
        \includegraphics[width=\linewidth, height=0.25\textheight]{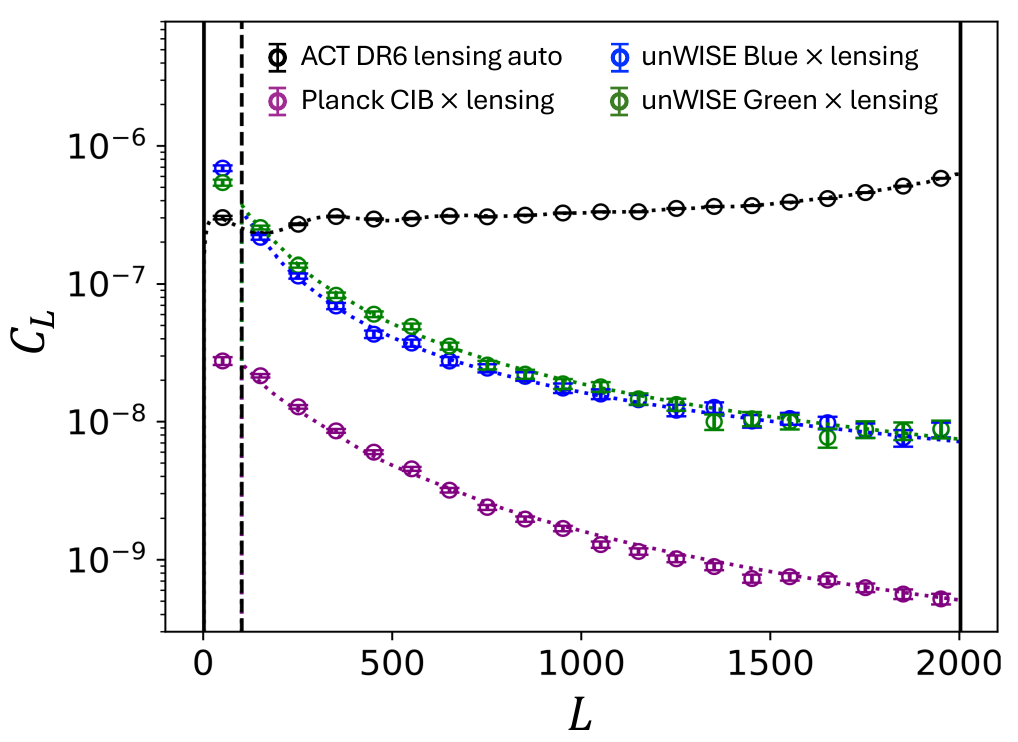}
    \end{minipage}%
    \hfill
    \begin{minipage}{0.49\textwidth}
        \centering
        \vspace{-1pt}
        \includegraphics[width=1.02\linewidth, height=0.255\textheight]{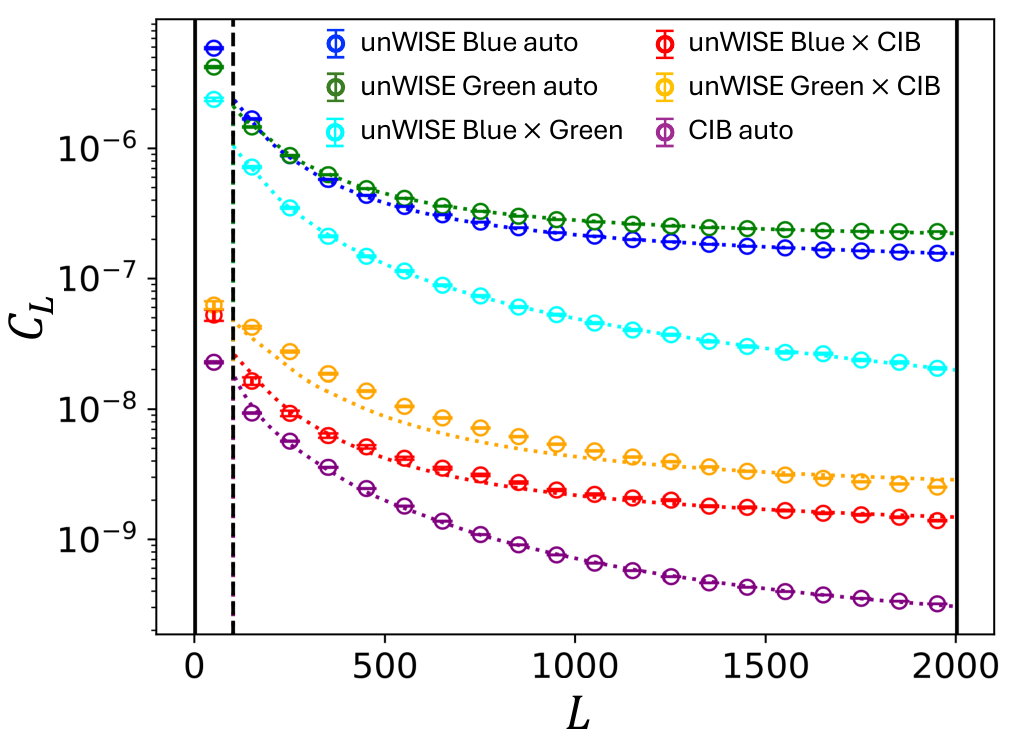}
    \end{minipage}
    \vspace{-5pt}
    \caption{\textit{Left}: auto-spectrum of the ACT DR6 MV lensing map and its cross-spectra with all external LSS tracers. \textit{Right}: auto- and cross-spectra between the unWISE Blue and Green samples and the Planck GNILC CIB map. Circles represent mode-decoupled bandpowers obtained with \texttt{NaMaster} using a bin width $\Delta L=100$, and dotted lines correspond to best-fit theoretical models from Eq.~\eqref{eq:limber}. The latter are a reasonable match to the data except for a slight deviation in the unWISE Green $\times$ CIB cross-spectrum, which was shown not to affect delensing results significantly. The CIB has units of $\text{MJy\,sr}^{-1}$; all other fields are dimensionless. Vertical lines denote the multipole ranges which enter our combined lensing estimator: $2\leq L\leq2000$ for CMB lensing (solid) and $102\leq L\leq2000$ for external tracers (dashed). Error bars, reflecting sample variance, are computed with the analytic approximation from Ref.~\cite{knox_1995}.}
    \label{fig:tracers_spectra}
\end{figure*}

As the QEs making up our internal lensing reconstructions each contain two powers of the masked CMB fields, we compute the mode-coupling matrix based on the square of the ACT mask for all spectra involving the ACT lensing map. While this does not fully capture the non-trivial way in which the lensing reconstruction pipeline convolves the mask with the signal, the approximation holds within statistical uncertainties~\cite{qu_2024}. Small residuals are mitigated by applying a multiplicative transfer function determined from simulations; explicitly, the correction to the cross-spectrum between ACT lensing and an arbitrary tracer $\hat{\kappa}^i$ is given by
\begin{equation}
    f^{\kappa}_{L}=\frac{\langle C_L^{\hat{\kappa}^{\rm{ACT}}\kappa}\rangle}{C_L^{\kappa\kappa}}.
\end{equation}
Here, the numerator is the \texttt{NaMaster} output for the ACT-like reconstruction $\hat{\kappa}^{\rm{ACT}}$ (associated with the square of the ACT mask) and the input convergence $\kappa$ (multiplied by the same mask as $\hat{\kappa}^i$). The denominator is the fiducial lensing power spectrum and angle brackets denote an ensemble average over simulations.

To avoid biases caused by statistical fluctuations in the measured correlations, we follow~\cite{yu_multitracer_2017} and fit smooth theoretical models (described in Sec.~\ref{subsec:lss_theory}) through the binned tracer cross-spectra before computing the weights in Eq.~\eqref{eq:kappa_comb}. For the CMB lensing power spectrum, we simply use $C_L^{\hat{\kappa}\hat{\kappa},\rm{th}}=C_L^{\kappa\kappa}+N_L^{\kappa\kappa}$, where the first term is the CAMB prediction for our fiducial cosmology and the reconstruction noise curve $N_L^{\kappa\kappa}$ is determined from simulations~\cite{qu_2024}. The observed power spectrum of the MV ACT lensing map and the corresponding model are displayed as the black circles and dotted line, respectively, in the left panel of Fig.~\ref{fig:tracers_spectra}, over the range $2\leq L \leq 2000$. As expected, $C_L^{\hat{\kappa}\hat{\kappa}}$ is signal-dominated on large scales and becomes noise-dominated for $L\geq250$. Conversely, correlations involving galaxies or the CIB are susceptible to large Galactic dust contamination at low $L$; we therefore restrict the fitting multipole range for external tracers to $L\geq 102$ in the next stages of the pipeline.

Assuming the cross-spectra of the unWISE overdensity maps $\hat{g}^i$ ($i\in\{1,2\}$) with ACT lensing to be a good proxy for their correlation with the true $\kappa$, we model these in the Limber approximation introduced in Sec.~\ref{subsec:lss_theory}. For each sample, we parametrize the bias evolution in Eq.~\eqref{eq:galaxy_kernel} as $b(z)=b_0^i+b_1^iz$, then use Eq.~\eqref{eq:limber} to fit for $b_0^i$ and $b_1^i$ by setting $C_L^{\hat{g}^i\hat{\kappa},\rm{th}}=C_L^{g^i\kappa}(b_0^i,b_1^i)$. Note that we only consider the range $102\leq L \leq 800$ for these fits as the linear bias assumption is expected to break down at large $L$. However, deviations appear to be small on the scales of interest: the blue and green theory curves in the left panel of Fig.~\ref{fig:tracers_spectra} remain a good approximation to the data up to $L=2000$.

It is important to mention at this stage that the purpose of our modelling choices for clustering statistics is only to find smooth functions matching the observed cross-spectra. Our model is not intended to extract any physical information from the inferred parameter values, so its simplistic nature is not a limitation for our analysis. For the same reason, we do not need to consider the cosmology scale cuts cited in~\cite{qu_2024} and~\cite{farren_atacama_2023}. A more detailed theoretical framework is presented in~\cite{farren_atacama_2023} to describe correlations between unWISE and ACT lensing.

With $b_0^i$ and $b_1^i$ fixed for both unWISE samples, we now fit their respective auto-spectra, which include a scale-independent shot noise component. Explicitly, we set $C_L^{\hat{g}^i\hat{g}^i,\rm{th}}=C_L^{g^ig^i}(b_0^i,b_1^i)+N^{g^ig^i}$, where the first term is obtained by evaluating Eq.~\eqref{eq:limber} for the relevant kernels and the noise term is inferred separately for each sample. Using the same form for the cross-spectrum between the Blue and Green galaxy-overdensity maps, we find the best-fit shot noise amplitude to be negligible (as expected for independent samples) and observe a slight mismatch between the data and the model. The discrepancy arises as a consequence of ignoring additional terms related to the lensing-magnification bias~\cite{farren_atacama_2023}, which are subdominant due to the strong clustering signal in the auto-spectra but become more significant when correlating different samples. We do not need to model these effects fully here as we are not concerned with physical interpretation; instead, we simply apply a multplicative correction expressed as a linear function of $l$ with two free parameters. As shown in the right panel of Fig.~\ref{fig:tracers_spectra}, this is sufficient to obtain an accurate fit to the data.

We follow a similar process for the CIB as we did for unWISE, first using its cross-spectrum with ACT lensing to determine the normalization factor $b_c$ in Eq.~\eqref{eq:cib_kernel}. This parameter is then fixed when fitting the auto-spectrum model, $C_L^{\hat{I}\hat{I},\rm{th}}=C_L^{II}(b_c)+N^{II}+A_dL^{\alpha_d}$, which includes the Limber approximation term, a constant shot-noise amplitude and a power-law term encoding dust contamination. The CIB auto-spectrum and cross-spectrum with lensing are displayed as purple lines in Fig.~\ref{fig:tracers_spectra}.

Finally, we fit the shot noise parameters in the cross-spectra between the unWISE galaxies and the CIB, $C_L^{\hat{g}^i\hat{I},\rm{th}}=C_L^{g^iI}(b_0^i,b_1^i,b_c)+N^{g^iI}$. Adding a dust contribution does not make any difference as the inferred amplitude is negligible. The red and yellow lines in the right panel of Fig.~\ref{fig:tracers_spectra} are a reasonable match to the data despite a slight deviation in the Green sample measurements. Possible delensing biases arising from such uncertainties in tracer spectra were investigated in~\cite{namikawa_simons_2022} and found not to have a significant impact on $\sigma(r)$ at SO's sensitivity level.

Using all aforementioned smooth theory curves, we then compute the coefficients from Eq.~\eqref{eq:kappa_comb} and construct the optimally weighted sum of tracers, keeping multipoles $2\leq L\leq 2000$ for the ACT lensing map and $102\leq L\leq 2000$ for the external tracers. 

The next stage of our analysis consists of building the template and characterizing its correlation with the true lensing $B$-modes in our sky. For this purpose, we extract $E$-modes from the ACT DR6 polarization map made of co-added data from the f90 and f150 bands. To downweight noise-dominated scales, we apply the diagonal Wiener filter shown in Eq.~\eqref{eq:wiener}, where $C_l^{EE}$ and $C_l^{EE}+N_l^{EE}$ are obtained from the average of 100 noise-free and 100 realistic ACT-like CMB simulations, respectively. Note that both sets of simulations are subjected to the Fourier-space filtering described in Sec.~\ref{section:ACT}.

To yield the lensing $B$-mode template, the Wiener-filtered $E$-modes are restricted to $100 \leq l \leq 2048$ and convolved with our optimal lensing convergence estimator $\hat{\kappa}^{\rm{comb}}$ as indicated in Eq.~\eqref{eq:lensing_B}. The lower bound of this multipole range is set to avoid scales at which $E$-modes are strongly suppressed by the $k$-space filter. For $\hat{\kappa}_{LM}^{\rm{comb}}$, we follow~\cite{qu_2024} and remove scales below $L=40$ where the mean-field bias surpasses the internally reconstructed lensing signal. 

The cross-spectrum between our template and the observed $B$-modes is computed by using \texttt{NaMaster} as explained at the beginning of this section, although the bin width is reduced to $\Delta l=40$ and additional precautions are taken to mitigate the mask-induced mixing of $E$- and $B$-modes. For polarized (spin-2) fields, the harmonic coefficients $a^w_{lm}$ in Eq.~\eqref{eq:pseudo-Cl} are replaced by the vector $\bm{P}^w_{lm}=\left(E^w_{lm},B^w_{lm}\right)^{T}$ and Eq.~\eqref{eq:namaster} becomes 
\begin{equation}\label{eq:namaster_pol}
    \left\langle\textrm{vec}\left(\hat{\mathbf{C}}^{ab,vw}_l\right)\right\rangle =\sum_{l'}{\mathbf{M}_{ll'}^{22}\textrm{vec}\left(\mathbf{C}_{l'}^{ab}\right)}.
\end{equation}
Here, $\textrm{vec}\left(\hat{\mathbf{C}}_{l}\right)=\left(C_l^{EE},C_l^{EB},C_l^{BE},C_l^{BB}\right)^T$ and the mode-coupling matrix $\mathbf{M}^{22}$ is $4\times 4$ for fixed multipoles $l$ and $l'$. Incomplete sky coverage therefore breaks the orthogonality of polarization; as a result, $E$-modes (which have significantly larger power in the primary CMB) leak into $B$-modes, increasing uncertainties on their power spectrum. \texttt{NaMaster} resolves this issue by incorporating $B$-mode purification, which projects out the contaminated part of the $B$-mode estimator~\cite{smith_pseudo-c_ell_2006} before inverting Eq.~\eqref{eq:namaster_pol}. We compute all cross-spectra between the template and observed $B$-modes over the ACT lensing mask and increase its apodization length to 10 degrees to facilitate the purification process.

Finally, the measured power spectra need to be corrected for the effects of the Fourier-space filter applied to the ACT maps. These manifest as a scale-dependent suppression of power, both in the $E$-modes entering the template and in the $B$-modes we seek to delens. The corresponding transfer functions $f_l^{EE}$ and $f_l^{BB}$ are evaluated from simulations, by computing the average cross-spectra $C_l^{EE_f}$ and $C_l^{BB_f}$ of 100 noise-free filtered ACT-like maps with their unfiltered counterparts before comparing to the underlying CAMB theory ($f_l^{EE}=\langle C_l^{EE_f}\rangle/C_l^{EE}$ and the same for $BB$). We then propagate the $E$-mode suppression to the template power spectrum $C_l^{B_tB_t,f}$ by substituting $C_l^{EE_f}$ into Eq.~\eqref{eq:Cl_tempxtemp}. This yields an estimated transfer function $f_l^{\rm{temp}}=C_l^{B_tB_t,f}/C_l^{B_tB_t}$, whose inverse is used to rescale the template auto-spectrum. Similarly, the cross-spectrum between the template and observed $B$-modes is divided by $f_l^{\rm{temp}}\times f_l^{BB}$. While this approximation only holds exactly in the limit of isotropic filtering, Appendix~\ref{appendix_A} explicitly verifies that the effects of the full ACT $k$-space filter are accurately captured.

\section{Results and discussion}\label{section:results}
\subsection{Multi-tracer lensing map}\label{section:results_kappa}

\begin{figure}[h]
    \centering
    \includegraphics[width=\linewidth, height=0.25\textheight]{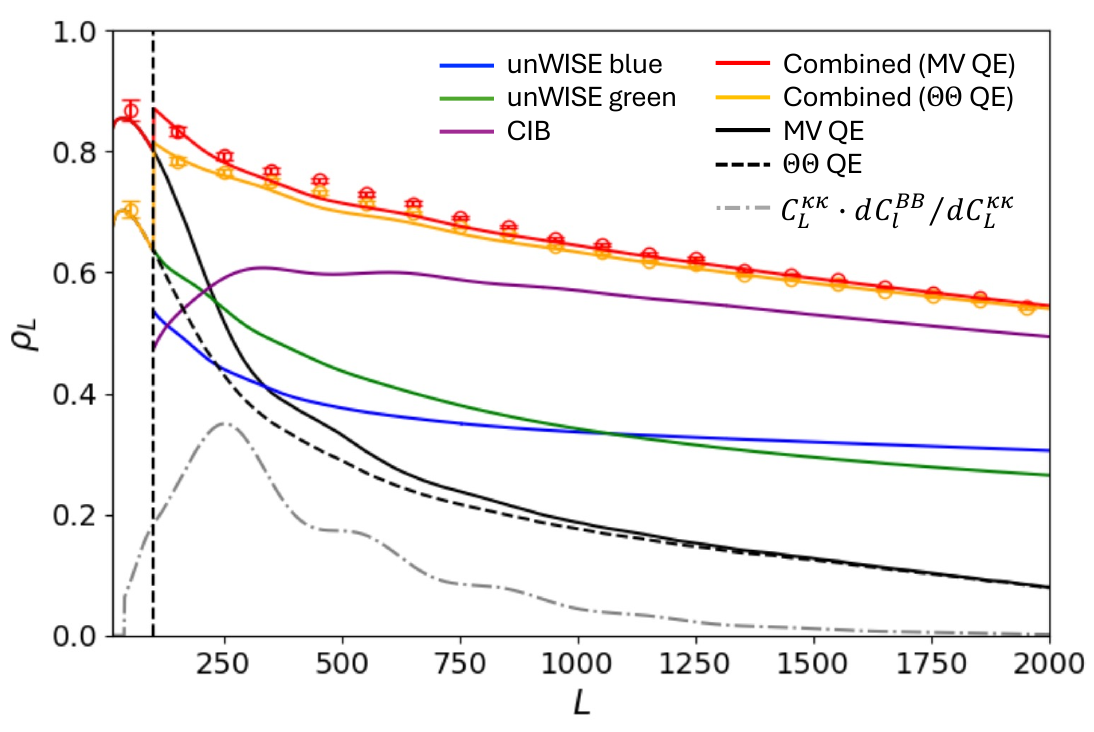}
    \vspace{-15pt}
    \caption{Correlation coefficients between the tracers used in this work and the true lensing convergence. Lines are obtained by evaluating Eq.~\eqref{eq:corr_coeff} for the best-fit theoretical models in Sec.~\ref{section:methodology}; the two unWISE samples and the CIB are displayed in blue, green and purple. The red and yellow curves represent the optimal combination of these tracers with the MV (black solid) and $\Theta$-only (black dashed) ACT DR6 internal reconstructions, respectively. The corresponding points are measured directly from the co-added $\hat{\kappa}$ maps using Eq.~\eqref{eq:kappa_comb_spectra}, with error bars obtained from the variance of 100 simulations. The discontinuity at $L=100$ appears as a result of discarding external tracers below this multipole. The grey dot-dashed line, given by $\sum_{30\leq l\leq 300}C_L^{\kappa\kappa} dC_l^{BB}/dC_L^{\kappa\kappa}$ and normalized for legibility, highlights the multipoles contributing the most to lensing $B$-modes on large scales.}
    \label{fig:rho_results}
\end{figure}

We start by presenting the results of the multi-tracer lensing reconstruction step of the pipeline. The correlation coefficients between each tracer and the true lensing convergence, estimated by applying Eq.~\eqref{eq:corr_coeff} to the best-fit theory cross-spectra from the previous section, are shown as continuous lines in Fig.~\ref{fig:rho_results}. As expected, the highest correlation at $L\leq250$ is achieved by the ACT DR6 MV internal reconstruction (black solid line) which peaks at approximately 85\%, a $10\%$ improvement compared to the Planck 2018 lensing map~\cite{planck_collaboration_planck_2020_lensing}. This value decreases to 70\% when only the $\Theta\Theta$ QE is considered (black dashed line). The CIB (purple) has the lowest $\rho_L$ at low $L$ as a result of dust contamination, but is up to 60\% correlated with the true $\kappa$ on the scales that contribute the most to the lensing $B$-modes ($250\leq L\leq 800$). The complementarity of these two tracers therefore allows their combination to remain an effective proxy of $\kappa$ over the full multipole range shown in Fig.~\ref{fig:rho_results}. Including the two unWISE samples, whose correlation coefficients with $\kappa$ vary between 60\% and 40\%, our co-added tracers (red and orange) achieve $55\% \leq \rho_L \leq 85\%$ and $55\% \leq \rho_L \leq 80\%$, respectively. Above $L\sim800$, the difference between the MV and $\Theta$-only cases becomes negligible as the CIB and unWISE contributions dominate. An extension of this result including additional galaxy samples from the DESI Legacy Imaging Surveys is presented in Appendix~\ref{appendix_C}.

While the true lensing convergence is not known, we can estimate $\rho_L^{\kappa\hat{\kappa}_{\rm{comb}}}$ by comparing the auto-spectrum of the combined tracer $\hat{\kappa}_{\rm{comb}}$ to the fiducial $C_l^{\kappa\kappa}$ from Sec.~\ref{section:methodology}; according to Eq.~\eqref{eq:kappa_comb_spectra}, the cross-spectrum $C_L^{\kappa\hat{\kappa}_{\rm{comb}}}$ is expected to be equal to $C_L^{\hat{\kappa}_{\rm{comb}}\hat{\kappa}_{\rm{comb}}}$. This method is applied to the MV and $\Theta$-only combined tracers as well as to our simulation suites, using the NaMaster pseudo-$C_l$ formalism in the overlap area between the ACT, unWISE and Planck masks. Results are shown as the red and yellow points, and are overall in good agreement with the theoretical prediction. The slight deviations at intermediate multipoles stem from small differences between the measured and fitted tracers cross-spectra in Fig.~\ref{fig:tracers_spectra} (in particular between the CIB and unWISE). Such uncertainties were investigated in~\cite{namikawa_simons_2022} and found not to have a significant impact on cosmological inference at SO's sensitivity level. In particular, combining LSS tracers with internal reconstructions (which dominate at low $L$) was shown to mitigate unmodelled lensing $B$-mode residuals susceptible to bias constraints on $r$.

\begin{figure}[h]
    \centering
    \includegraphics[width=\linewidth, height=0.24\textheight]{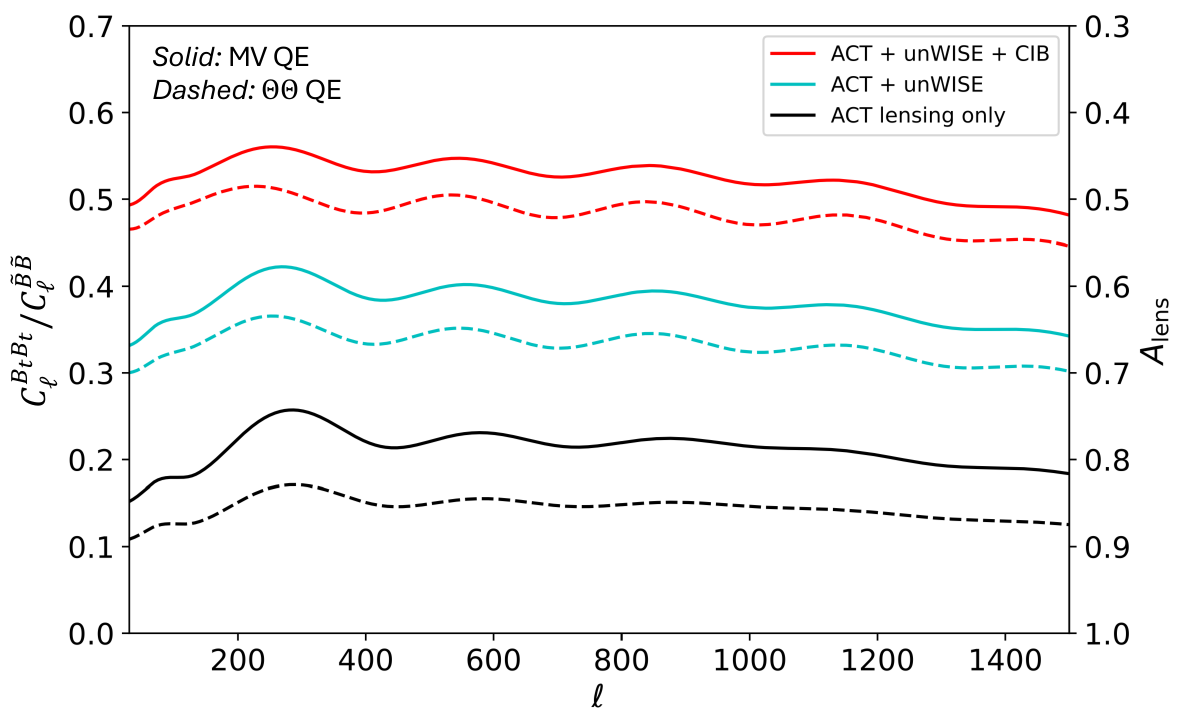}
    \vspace{-15pt}
    \caption{Maximum delensing efficiency $C_l^{B_tB_t}/C_l^{BB}$ and corresponding $A_{\text{lens}}$ achievable for different combinations of tracers, obtained by evaluating Eq.~\eqref{eq:Cl_tempxtemp} assuming ideal $E$-modes. Solid and dashed lines represent combinations including the MV and $\Theta$-only ACT lensing maps, respectively. While the MV internal reconstruction on its own can only remove up to 21\% of the lensing $B$-modes ($A_{\text{lens}}\approx 0.79$), adding the unWISE samples and the CIB allows us to reach delensing efficiencies of 38\% and 52\% ($A_{\text{lens}}\approx 0.62$ and $0.48$); this clearly illustrates the benefits of our multi-tracer approach.}
    \label{fig:Alens}
\end{figure}

Figure~\ref{fig:Alens} illustrates the maximal delensing efficiency achievable by various combinations of mass tracers, assuming a template built with ideal $E$-modes. This quantity is defined as the ratio between $C_l^{B_tB_t}$ (obtained by substituting Eq.~\eqref{eq:kappa_comb_spectra} into Eq.~\eqref{eq:Cl_tempxtemp} for the tracer in question) and the lensing $B$-mode power spectrum in our fiducial cosmology. Its average over $l$ corresponds to $1-A_{\textrm{lens}}$, where $A_{\textrm{lens}}$ (defined in Eq.~\ref{eq:Alens}) parametrizes the fractional $B$-mode residual power after map-level delensing.

On its own, the MV internal reconstruction (solid black line) reaches a mean delensing efficiency of 21\%; this value is consistent with previous results using ACT DR4 data~\cite{han_atacama_2020} and represents a significant increase compared to the 7\% achieved by the Planck lensing map~\cite{carron_internal_2017}. Internal delensing will continue to gain importance in the next few years; for example, it is expected to remove up to 40\% of the lensing $B$-mode power by the end of SO's initial survey~\cite{the_simons_observatory_collaboration_simons_2019,namikawa_simons_2022}, and could reach a 90\% efficiency with the proposed CMB-HD experiment~\cite{MacInnis2024_CMBHD}.

Despite these improvements in the quality of internal lensing reconstructions, external LSS tracers currently provide most of the power for $B$-mode delensing and will remain essential as SO accumulates its first few years of data. Adding the unWISE galaxy samples to the ACT DR6 QE (cyan line in Fig.~\ref{fig:Alens}) nearly doubles the achievable efficiency, bringing it to an average of 38\%. With the CIB also included, our combined tracer is able to remove up to 52\% of the lensing $B$-mode power (red line). Using the $\Theta\Theta$ internal reconstruction instead of its MV counterpart results in a moderate loss of delensing power, with predicted efficiencies ranging from 14\% (QE only, dashed black line) to 48\% (multi-tracer case, dashed red line).

In~\cite{yu_multitracer_2017}, a co-addition of the Planck lensing and GNILC CIB maps with WISE galaxies was found to reach a maximum delensing fraction of 43\%. The higher values obtained here are explained by the improvement in sensitivity between Planck and ACT as well as by the increased depth of unWISE compared to the original WISE catalog.

\subsection{Delensing performance}\label{section:results_del}

\begin{figure*}[!t]
    \centering
    \begin{minipage}{.49\textwidth}
        \centering
        \includegraphics[width=\linewidth, height=0.23\textheight]{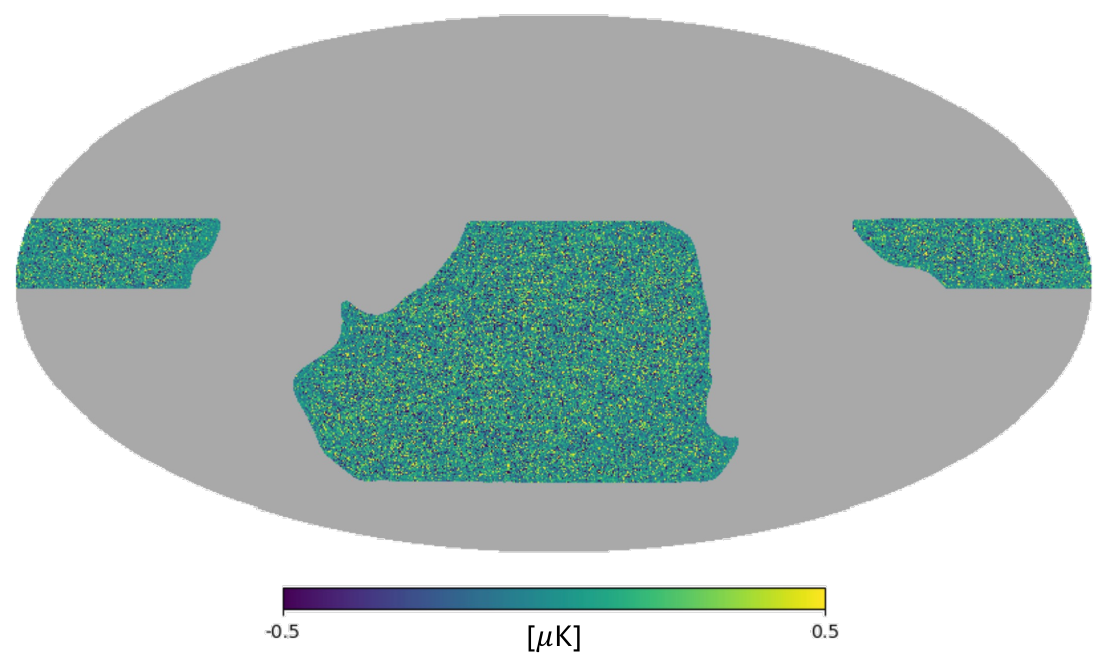}
    \end{minipage}%
    \hfill
    \begin{minipage}{0.49\textwidth}
        \centering
        \includegraphics[width=\linewidth, height=0.228\textheight]{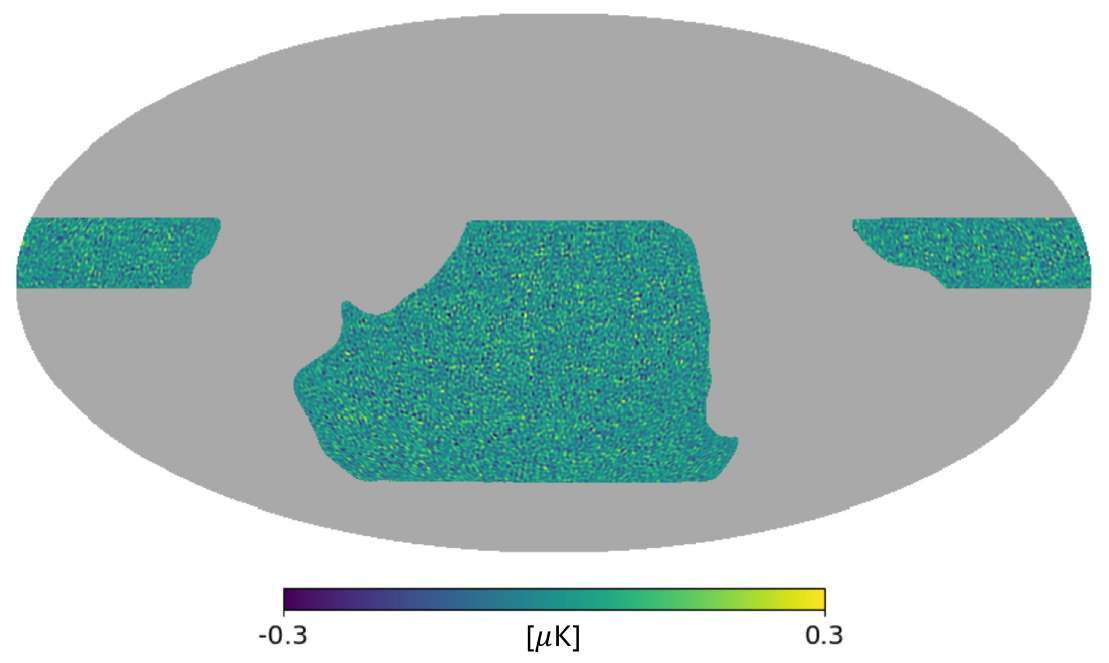}
    \end{minipage}
    \vspace{-5pt}
    \caption{Lensing $B$-mode template constructed by convolving our combined lensing tracer (including the MV internal reconstruction) with ACT DR6 $E$-modes, and projected as a scalar field onto the ACT lensing mask. The left panel shows the full range of scales while the right panel displays a low-pass-filtered version with $l \leq 300$.}
    \label{fig:template_maps}
\end{figure*}

\begin{figure*}[!t]
    \centering
    \begin{minipage}{.49\textwidth}
        \centering
        \includegraphics[width=\linewidth, height=0.25\textheight]{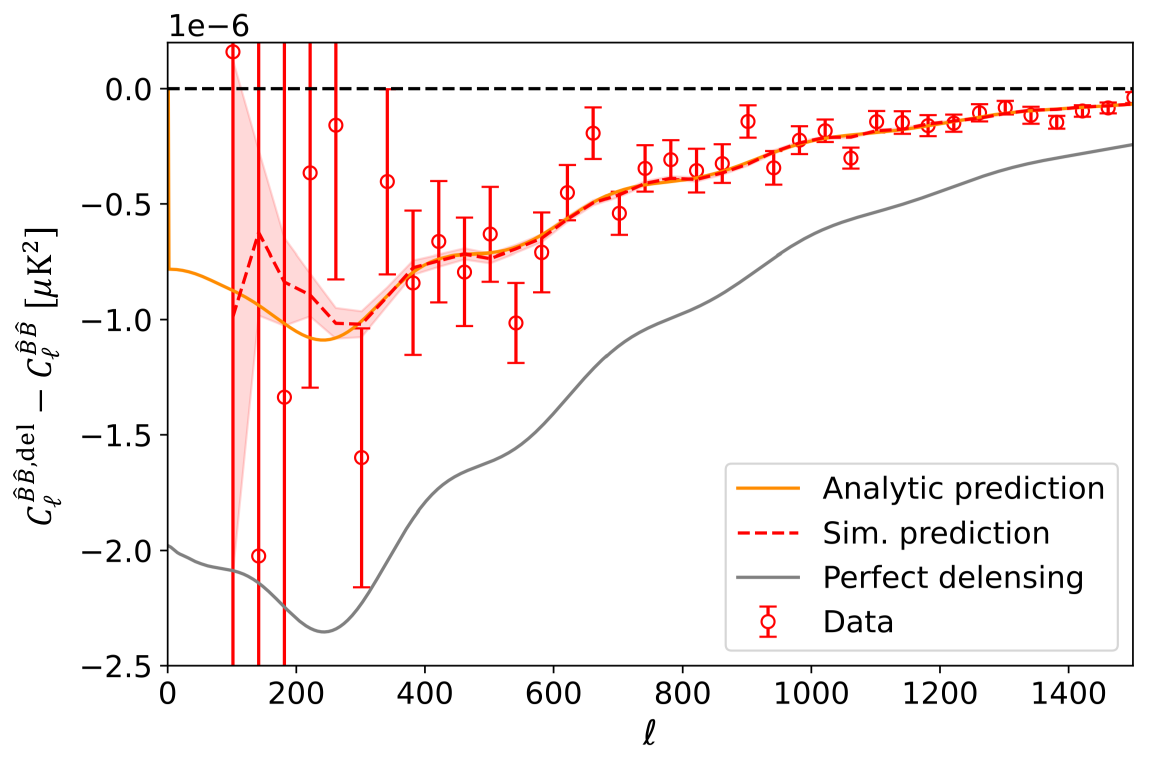}
    \end{minipage}%
    \hfill
    \begin{minipage}{0.49\textwidth}
        \centering
        \includegraphics[width=\linewidth, height=0.25\textheight]{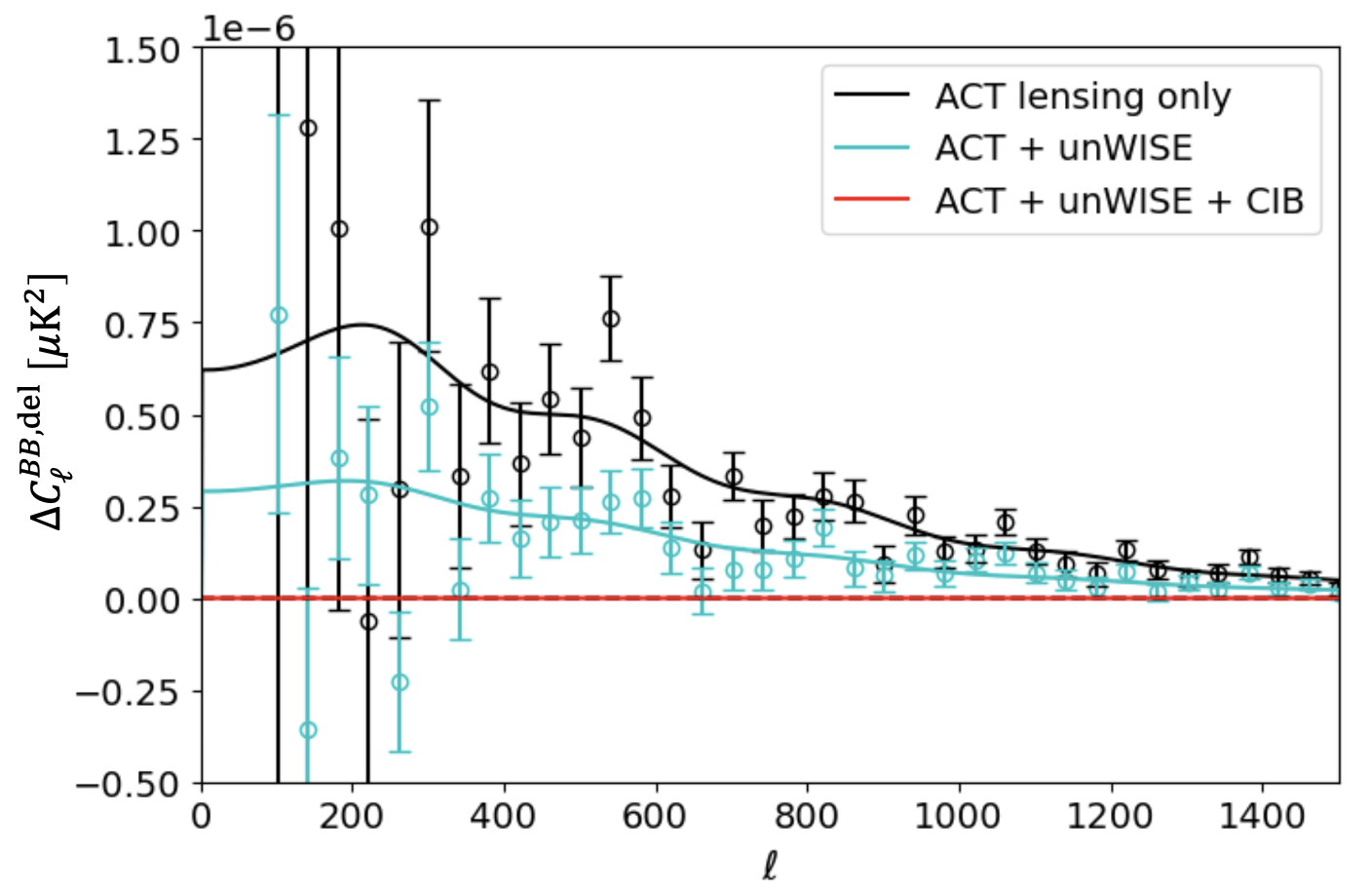}
    \end{minipage}
    \vspace{-5pt}
    \caption{\textit{Left}: noise-cancelling difference between the delensed and observed ACT $B$-mode power spectra (Eq.~\ref{eq:delta_BB}). Our template is built using the $\Theta$-only ACT DR6 lensing map, the CIB and unWISE. Points are obtained by computing the transfer-function-corrected template auto- and cross-spectra $C_l^{B_tB_t}$ and $C_l^{\hat{B}B_t}$. The $1\sigma$ error bars are estimated from the variance of $\Delta C_l^{BB}$ over 100 simulations, for which we found the diagonal covariance to be a reasonable approximation. The grey line corresponds to perfect delensing, while the yellow line shows the analytical prediction from Eqs~\eqref{eq:Cl_Bxtemp} and~\eqref{eq:Cl_tempxtemp}. The mean of 100 simulations appears as the dashed red line, with the corresponding $1\sigma$ standard error as the shaded area. \textit{Right}: excess residual lensing $B$-mode power observed when building our $\kappa$ map from restricted subsets of tracers, compared to the baseline case (all combined) in red. This is given by $C_l^{BB,\rm{del}}(\hat{\kappa}_{i})-C_l^{BB,\rm{del}}(\hat{\kappa}_{\rm{comb}})$, for $\hat{\kappa}_i$ constructed from ACT+unWISE (cyan) and ACT only (black). Again, continuous lines are analytical predictions, points are measured from the data and error bars are obtained from simulations.}
    \label{fig:del_performance_ACT}
\end{figure*}

Using the co-added tracers characterized in the previous section, we now assess the performance of the lensing $B$-mode template built as indicated at the end of Sec.~\ref{section:methodology}.

The maps in Fig.~\ref{fig:template_maps} show two versions of our template directly projected onto the ACT DR6 unmasked sky area. While the left panel displays the full multipole range $l\leq2048$, the right panel is restricted to $l\leq 300$ to focus on the scales most relevant for primordial $B$-mode searches. Both maps are consistent in amplitude throughout the observed sky patch, taper off as expected near the edges of the mask and do not contain any visible foreground or noise artifacts.

To demonstrate delensing on current CMB data, we quantify the decrease in lensing $B$-mode power following the subtraction of our template from the observed maps. The delensed $B$-modes, given by $\hat{B}_{lm}-B^t_{lm}=\tilde{B}_{lm}+n_{lm}-B^t_{lm}$ (where $n_{lm}$ represents an uncorrelated noise contribution), have the following power spectrum: $C_l^{\hat{B}\hat{B}, \rm{del}}=C_l^{\tilde{B}\tilde{B}}+N_l-2C_l^{\hat{B}B_t}+C_l^{B_tB_t}$, where $N_l$ is the noise power spectrum. For ACT and Planck, both this quantity and the total measured $B$-mode power spectrum $C_l^{\hat{B}\hat{B}}=C_l^{\tilde{B}\tilde{B}}+N_l$ are noise-dominated. However, taking the difference 
\begin{equation}\label{eq:delta_BB}
    \Delta C_l^{BB} = C_l^{\hat{B}\hat{B},\textrm{del}}-C_l^{\hat{B}\hat{B}}= C_l^{B_tB_t}-2C_l^{\hat{B}B_t}
\end{equation}
cancels out the noise term and allows us to detect the effect of delensing at high levels of significance.

Our measurement of $\Delta C_l^{BB}$ for ACT DR6 $B$-modes is shown in the left panel of Fig.~\ref{fig:del_performance_ACT}.
We evaluate this quantity by computing the power spectrum of the lensing template ($C_l^{B_tB_t}$) as well as its cross-spectrum $C_l^{\hat{B}B_t}$ with observed $B$-modes over the apodized ACT mask. The purpose of this result is to validate our template on intermediate scales; we therefore use the $\Theta$-only internal lensing reconstruction in our combined tracer to avoid overlap between the $EB$ QE and the $B$-modes to delens. As shown in Appendix~\ref{appendix_B} and~\cite{Sehgal2017InternalDelensing,lizancos_impact_2021}, correlations resulting from such overlap would significantly bias $C_l^{\hat{B}B_t}$. While this bias is dominated by correlated noise at current sensitivity levels, the signal contribution will also be important for future experiments. Note that this bias would not be an issue if our only goal was to delens large-scale $B$-modes: in this case, overlap could be avoided with a simple multipole cut (as will be discussed below).

An analytical model for $\Delta C_l^{BB}$ can be obtained by substituting the $E$-mode Wiener filter estimated from simulations and the relevant $\rho_L$ curve from Fig.~\ref{fig:rho_results} into Eqs~\eqref{eq:Cl_Bxtemp} and~\eqref{eq:Cl_tempxtemp}. Within statistical uncertainties (determined from the variance of $\Delta C_l^{BB}$ over 100 ACT-like simulations~\cite{atkins_atacama_2023}), the points in Fig.~\ref{fig:del_performance_ACT} are consistent with this theoretical prediction (yellow line). 
Scaling the model by an amplitude $A_{\Delta C_l}$ and using a diagonal covariance matrix to encode statistical errors, we obtain a best-fit value 
\begin{equation}\label{eq:amp_ACT}
    A_{\Delta C_l}=0.95\pm0.05 \quad (1\sigma)
\end{equation}
in the multipole range $100\leq l \leq 1500$. This result is within 1$\sigma$ of our expectation ($A_{\Delta C_l}=1$) and indicates that delensing is detected with a signal-to-noise ratio (SNR) of 19.7. The mean delensing efficiency of our template is 39\%, which represents a 19\% decrease compared to the 48\% achievable with ideal $E$-modes (see Sec.~\ref{section:results_kappa}). Improving $E$-mode quality with the SO LAT will therefore contribute to better delensing performance; however, while polarization noise was the main limitation for previous analyses using Planck $E$-modes~\cite{planck_collaboration_planck_2020_lensing}, the fidelity of the lensing map is now the most important factor.

The dashed red line in Fig.~\ref{fig:del_performance_ACT} corresponds to the average $\Delta C_l^{BB}$ obtained for our set of 100 simulated templates and ACT-like polarization maps, with the shaded area showing the 1$\sigma$ error on the mean. This result is in excellent agreement with the theoretical prediction at $l\gtrsim 300$, indicating that our measurements on intermediate and small scales are robust to systematics induced by masking, anisotropic filtering, and noise inhomogeneities. Such non-idealities, which are not captured by the theoretical model but are replicated in our simulations, may partly contribute to the deviation observed at $l<300$. We therefore always consider the average of simulations to be more representative of the data than analytical approximations. The high level of scatter at low $l$ is expected as ACT's design and scanning strategy are optimized to produce high-resolution maps rather than to target large-scale $B$-modes.

In the right panel of Fig.~\ref{fig:del_performance_ACT}, we investigate the loss of delensing efficiency following the removal of external tracers from our combined lensing reconstruction. This is illustrated by plotting the increase in residual lensing $B$-mode power, $C_l^{BB,\rm{del}}(\hat{\kappa}_{i})-C_l^{BB,\rm{del}}(\hat{\kappa}_{\rm{comb}})$, observed when delensing the ACT DR6 polarization map with a subset of tracers $\hat{\kappa}_{i}$ instead of the full co-added set $\hat{\kappa}_{\rm{comb}}$. Keeping the $\Theta\Theta$ internal reconstruction and the two unWISE samples but removing the CIB (cyan), we detect excess residuals at the level of $0.12C_l^{BB}$, meaning that the delensing efficiency decreases from 39\% to 27\%. With the ACT lensing map only (black), the template captures 12\% of the total lensing $B$-mode power. 

In both cases, the observed excess residuals are consistent with theoretical expectations: comparing our data points to the cyan and black curves in Fig.~\ref{fig:del_performance_ACT}, we obtain reduced $\chi^2$ values of 1.2 and 1.3, respectively. With 35 degrees of freedom, this corresponds to probabilities to exceed (PTEs) of 0.16 and 0.09, which are within the acceptable range. The agreement between model and measurements, both with and without including the CIB in our template, indicates that biases related to foreground contamination are small within statistical uncertainties. Possible issues arising from correlated dust residuals in the CIB and CMB data were studied extensively in~\cite{lizancos_delensing_2022}; filtering out the largest scales ($L<100$) and masking the Galactic plane in the CIB map prior to building the lensing template were shown to reduce the impact of dust on the power spectrum of delensed $B$-modes. Furthermore, the multitracer approach implemented in the present work leads to smaller foreground-related biases than using the CIB as the sole mass tracer. 

\subsubsection*{Large-scale $B$-mode delensing}

Although the results shown in Fig.~\ref{fig:del_performance_ACT} successfully validate our pipeline on intermediate and small scales, ACT $B$-modes at $l\leq400$ are increasingly noisy and down-weighted by the Fourier-space filter, becoming inaccessible below $l\sim100$. To assess the performance of our template on the large scales most relevant for PGW searches, we delens the publicly available Planck PR3 SMICA map\footnote{The PR3 Legacy CMB maps can be downloaded at \url{https://esdcdoi.esac.esa.int/doi/html/data/astronomy/planck/CMB_Maps.html}.}. SMICA (Spectral Matching Independent Component Analysis) is a non-parametric component separation method which linearly combines data from the nine Planck frequency channels (between 30 and 857~GHz) to minimize foreground residuals~\cite{planck_collaboration_smica}. The resulting polarization maps have an effective Gaussian beam of 5 arcmin.

\begin{figure}[h]
    \centering
    \includegraphics[width=\linewidth, height=0.25\textheight]{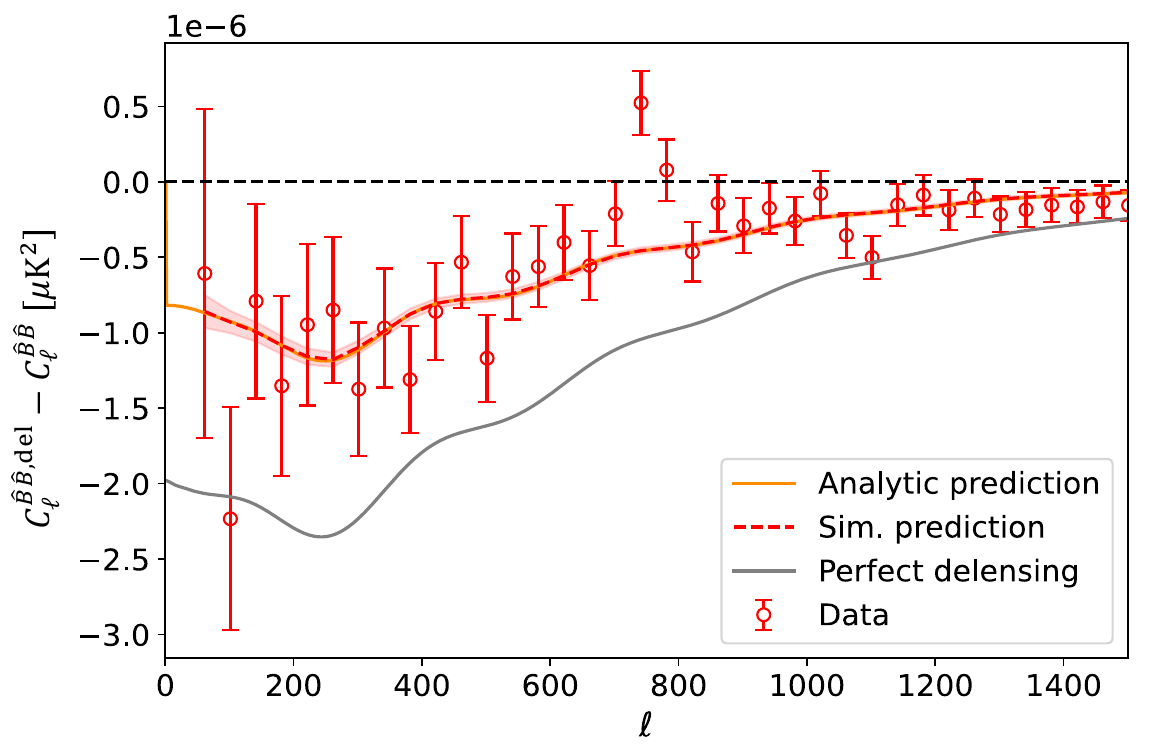}
    \vspace{-15pt}
    \caption{Same as left panel of Fig.~\ref{fig:del_performance_ACT}, but with $C_l^{\hat{B}B_t}$ computed by cross-correlating our template with Planck SMICA $B$-modes. The co-added lensing tracer used here includes the MV ACT internal reconstruction. The dashed line is obtained from the cross-spectra of our simulated templates with the true lensing $B$-modes in the input CMB maps. Error bars reflect cosmic variance.}
    \label{fig:del_performance_Planck}
\end{figure}

In Fig.~\ref{fig:del_performance_Planck}, we plot the lensing $B$-mode power difference $\Delta C_l^{BB}$ defined in Eq.~\eqref{eq:delta_BB}, with $C_l^{\hat{B}B_t}$ corresponding to the cross-spectrum between our template and Planck $B$-modes (still measured within the ACT DR6 mask). This time, our multi-tracer lensing reconstruction includes the MV QE from ACT, which uses CMB data in the range $600\leq l \leq 3000$. In the absence of correlated noise between the ACT and Planck CMB maps, the internal bias induced by mode overlap at $l>600$ (see Appendix~\ref{appendix_B}) becomes negligible within statistical uncertainties; our choice of QE is further justified as the main purpose of Fig.~\ref{fig:del_performance_Planck} is to characterize our template on large scales ($l\leq300$), which remain unaffected.

Our $\Delta C_l^{BB}$ measurements are generally consistent with the 42\% mean delensing efficiency predicted by Eqs~\eqref{eq:Cl_Bxtemp} and~\eqref{eq:Cl_tempxtemp} (yellow line), and exhibit significantly less scatter than in Fig.~\ref{fig:del_performance_ACT} at low $l$. This is consistent with expectations: indeed, ACT has higher noise levels than Planck on large scales due to atmospheric contamination. As a sanity check, we also plot the mean $\Delta C_l^{BB}$ obtained by cross-correlating simulated templates with the input lensing $B$-modes used to generate the associated mock CMB maps. Assuming uncorrelated noise and no $k$-space filtering for Planck, this result should be compatible with our observations. The dashed red line in Fig.~\ref{fig:del_performance_Planck} indeed clearly agrees with the points and with the idealized analytical prediction, implying negligible impact from masking and noise anisotropies.

Over the full range of scales shown in Fig.~\ref{fig:del_performance_Planck}, we infer the amplitude
\begin{equation}\label{eq:amp_planck_full}
    A_{\Delta C_l}\big|_{30\leq l\leq1500}=0.87\pm0.09,
\end{equation}
corresponding to a 10$\sigma$ detection significance. This value is within 1.5$\sigma$ of $A_{\Delta C_l}=1$, indicating a reasonable agreement with theoretical expectations; however, it may be biased slightly low by the outlier at $l\sim750$ and by the mode-overlap bias above $l>600$, driven by the shared lensing $B$-modes. The outlier was not observed in the ACT delensing results (Fig.~\ref{fig:del_performance_ACT}) nor in the template auto-spectrum (see Fig.~\ref{fig:temp_cross_auto} later), which suggests that the issue does not originate from the template itself. Further tests were carried out using different subsets of mass tracers (ACT DR6 internal reconstruction only, external tracers only, ACT + unWISE without the CIB): the outlier appeared in all cases when cross-correlating the template with Planck $B$-modes but did not affect any other results. We therefore conclude that this point is probably an artifact of the Planck data and is not cause for concern. Note that the Planck $BB$ power spectrum is noise-dominated and does not contain any statistically significant outliers, preventing us from directly confirming this interpretation.

Restricting our measurement to the scales of interest for PGW ($30\leq l \leq 300$), the best-fit amplitude
\begin{equation}\label{eq:amp_planck_restr}
    A_{\Delta C_l}\big|_{30\leq l\leq300}=0.99\pm0.23
\end{equation}
is in excellent agreement with theory. Our template achieves a 47\% reduction of lensing $B$-mode power in this multipole range, which we detect at 4.3$\sigma$. These findings can be compared to the results of~\cite{planck_collaboration_planck_2020_lensing}, where a template using $E$-modes and QEs from Planck in addition to the 545~GHz GNILC CIB captured 21.7\% of the low-$l$ lensing $B$-modes. On smaller scales, internal delensing with ACT DR4 data led to a 20\% reduction in lensing $B$-mode power~\cite{han_atacama_2020}. The highest delensing efficiency achieved prior to the present paper (28\% in the range $300\leq l \leq 2300$) was reported in~\cite{manzotti_cmb_2017}; although this work used a \textit{Herschel} CIB map as the sole LSS tracer, high-resolution $E$-modes from SPTpol allowed for increased delensing power compared to Planck.

The results presented here therefore constitute a significant step forward relative to previous delensing studies. Further improvements will be forthcoming in the next five years, with the SO LAT and SPT-3G~\cite{sobrin_design_2022} expected to provide $E$-modes and internal lensing maps of unprecedented quality. The upcoming Rubin LSST~\cite{ivezic_lsst_2018} will also produce galaxy samples whose correlation with the true $\kappa$ should surpass that of the CIB, bringing the forecasted delensing efficiency to around $65\%$ by the end of SO's nominal mission~\cite{the_simons_observatory_collaboration_simons_2024}.

\subsection{Consistency tests}\label{sec:results_tests}

We conclude our investigation by performing additional consistency checks to verify the robustness of our results to various systematic effects. 

The first test aims to assess whether our combined estimator $\hat{\kappa}^{\rm{comb}}$ is correctly normalized and whether its correlation with the true lensing convergence is consistent with theoretical predictions. Although we do not have access to the true $\kappa$, we assume the form $\hat{\kappa}^{\rm{ACT}}_{LM}=\kappa_{LM}+n_{LM}$ for the ACT lensing map, with $n_{LM}$ being uncorrelated noise. Setting $\hat{\kappa}^0=\hat{\kappa}^{\rm{ACT}}$ in Eq.~\eqref{eq:kappa_comb}, we then expect
\begin{multline}\label{eq:kappa_consistency}
    C_L^{\hat{\kappa}^{\rm{comb}}\hat{\kappa}^{\rm{ACT}}}=\sum_{ij}\left(\bm{C}_L^{-1}\right)^{ij}C_L^{j\kappa}C_L^{i0} \\
    = \sum_{ij}\left(\bm{C}_L^{-1}\bm{C}_L\right)^{j0}C_L^{j\kappa}=C_L^{0\kappa} = C_L^{\kappa\kappa},
\end{multline}
where $C_L^{\kappa\kappa}$ is the fiducial lensing power spectrum.

\begin{figure}[h]
    \centering
    \includegraphics[width=\linewidth, height=0.25\textheight]{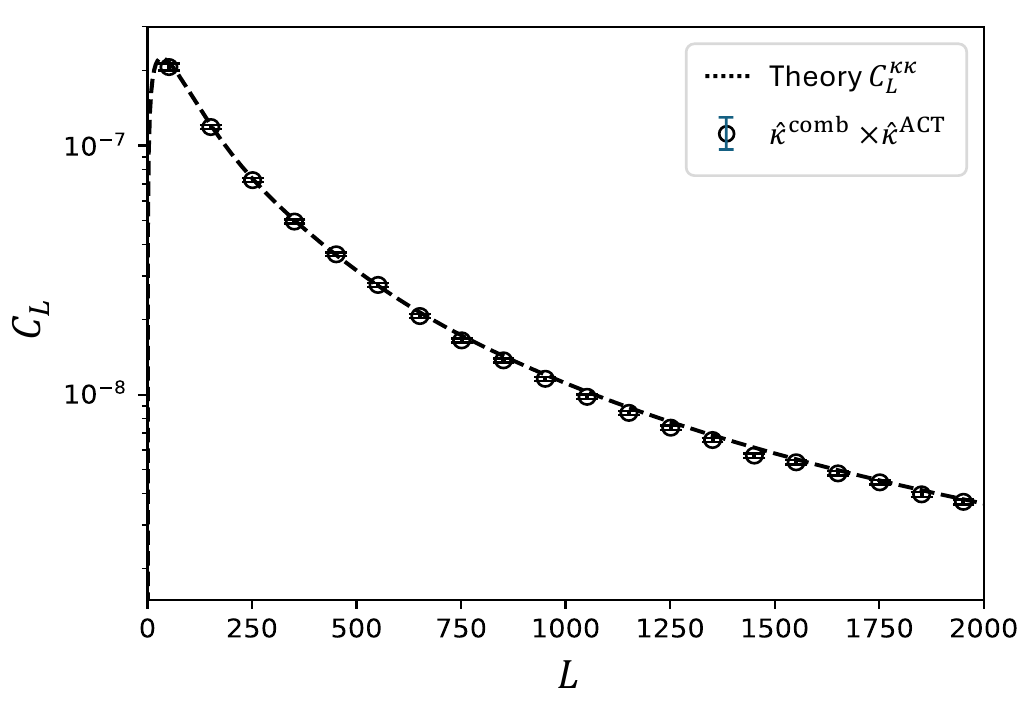}
    \vspace{-15pt}
    \caption{Cross-spectrum between our combined lensing tracer and the MV ACT lensing map. Error bars are obtained from the variance of 100 simulations. As a consequence of the optimal weighting in Eq.~\eqref{eq:kappa_comb}, this quantity is expected to be equal to the fiducial $C_l^{\kappa\kappa}$, shown as the dashed line.}
    \label{fig:kcomb_cross_kACT}
\end{figure}

The cross-spectrum measured between our multi-tracer $\kappa$ map and the ACT MV internal reconstruction is compared with the theoretical $C_L^{\kappa\kappa}$ from CAMB in Fig.~\ref{fig:kcomb_cross_kACT}, showing excellent agreement on the scales most relevant for lensing $B$-modes. Small deviations, attributable to modelling uncertainties in the tracers cross-spectra, remain within statistical errors. This indicates that our combined lensing estimator is robust to such residual systematics and that the map-level co-addition of tracers does not induce any spurious correlations.

\begin{figure}[h]
    \centering
    \includegraphics[width=\linewidth, height=0.25\textheight]{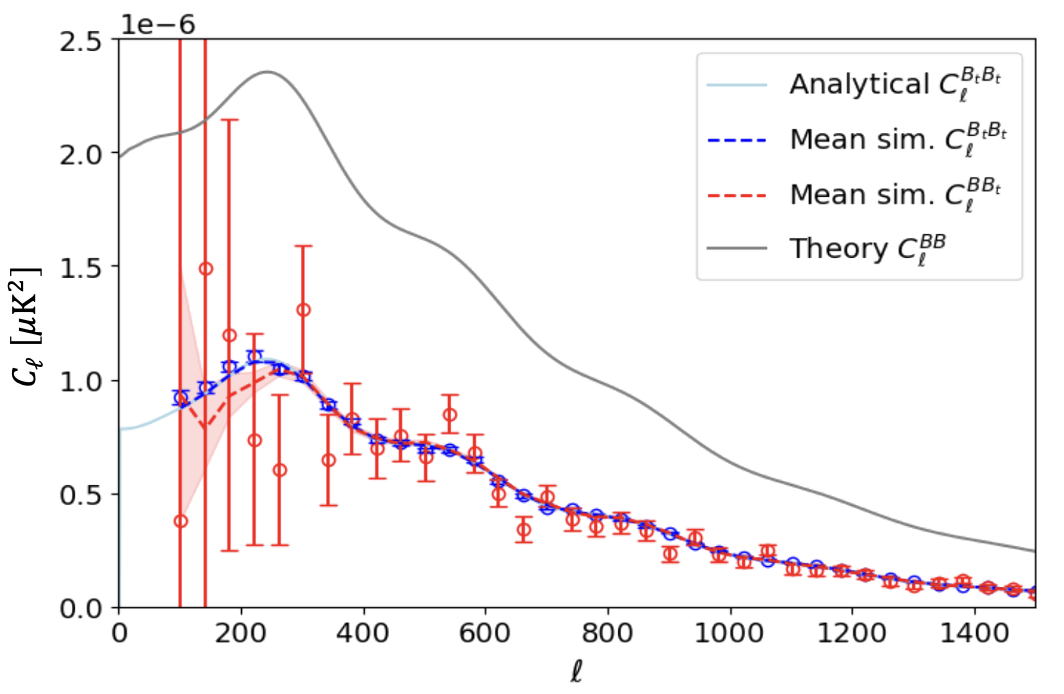}
    \vspace{-15pt}
    \caption{Template auto-spectrum $C_l^{B_tB_t}$ (blue) and cross-spectrum with ACT DR6 $B$-modes $C_l^{\hat{B}B_t}$ (red), corrected for their respective transfer functions. Our template is built using the $\Theta$-only ACT DR6 lensing map, the CIB and unWISE. The solid cyan line shows the analytical prediction from Eq.~\eqref{eq:Cl_Bxtemp}, which is identical to Eq.~\eqref{eq:Cl_tempxtemp}. The means of $C_l^{B_tB_t}$ and $C_l^{\hat{B}B_t}$ over 100 simulations appear as the dashed blue and red lines, respectively. The $1\sigma$ uncertainties on the points (error bars) and on the mean (shaded area) are obtained from the variance of these simulations.}
    \label{fig:temp_cross_auto}
\end{figure}

The second test consists of comparing the auto-spectrum of our template to its cross-spectrum with observed $B$-modes, to verify the equivalence of Eqs~\eqref{eq:Cl_Bxtemp} and~\eqref{eq:Cl_tempxtemp}. In Fig.~\ref{fig:temp_cross_auto} (where we used the $\Theta\Theta$ QE and ACT $B$-modes), the blue points corresponding to $C_l^{B_tB_t}$ closely follow the analytical expression represented by the solid cyan line as well as the mean of 100 simulations shown in dashed blue. While statistical errors on $C_l^{\hat{B}B_t}$ (red points) are significantly larger, the results are consistent with expectations: scaling the predicted cross-spectrum by an amplitude $A_{\times}$ (with $A_{\times}=1$ meaning $C_l^{\hat{B}B_t}=C_l^{B_tB_t}$), we obtain the best-fit value
\begin{equation}\label{eq:amp_planck_restr}
    A_{\times}=0.97\pm0.02,
\end{equation}
and a reduced $\chi^2$ of 1.1 with 34 degrees of freedom (corresponding to a PTE of 0.32). The red dashed line computed from simulations is also in excellent agreement with Eq.~\eqref{eq:Cl_Bxtemp} for $l\gtrsim300$ and remains within uncertainties on larger scales. Qualitatively similar results were obtained for the cross-spectrum between our template and Planck $B$-modes.

Besides confirming that survey non-idealities such as inhomogeneous noise, masking, and filtering only marginally affect our measurements, this test is particularly useful to rule out significant foreground biases. Indeed, correlated dust residuals in the CIB ($I$), the $E$-modes making up the template, and the $B$-modes to delens contribute higher-point functions $\langle EIEI\rangle_c$ and $\langle BEI\rangle_c$ to $C_l^{B_tB_t}$ and $C_l^{\hat{B}B_t}$,  respectively~\cite{lizancos_delensing_2022}. These bias terms impact the template auto- and cross-spectrum differently, leading to deviations from equivalence and departures from theoretical predictions. The same is true for possible mischaracterizations in the redshift distribution of galaxies~\cite{baleato_lizancos_impact_2023}. The absence of such effects in Fig.~\ref{fig:temp_cross_auto} indicates that Galactic dust is sufficiently mitigated by our masking schemes and multipole cuts, and that the impact of anisotropic galaxy redshift distributions is negligible at this stage. As mentioned in Sec.~\ref{section:results_del}, more stringent foreground cleaning measures will be implemented when delensing upcoming SO data with lower noise levels on large scales.

\section{Conclusions and outlook}\label{section:conclusion}

In this paper, we characterized a lensing $B$-mode template built by optimally co-adding an internal lensing reconstruction from ACT DR6, galaxy-overdensity maps based on the unWISE Blue and Green samples, and a CIB map from Planck. This work is motivated by the increasing importance of delensing for SO and next-generation CMB experiments aiming to set tight constraints on the tensor-to-scalar ratio.

Our combined lensing map was shown to be 55--85\% correlated with the true lensing convergence $\kappa$ and to reach a maximum potential delensing efficiency of 52\% in the range $30\leq l \leq 1500$, over 23\% of the sky. We also illustrated the benefits of the multi-tracer approach by constructing variants based on different subsets of tracers and quantifying the resulting loss of delensing power.

The lensing $B$-mode templates obtained by convolving two versions of our co-added $\kappa$ estimator with ACT DR6 $E$-modes were then validated on existing CMB maps. Restricting the internal reconstruction to the $\Theta\Theta$ QE (to avoid overlap between ACT $B$-modes in the template and those we aim to delens), we first demonstrated cross-spectral delensing on ACT $B$-modes and observed a reduction in power by approximately 39\% at $100\leq l \leq 1500$. Using the MV combination of QEs, we repeated the process with the Planck SMICA map and removed about 47\% of the lensing power ($A_{\text{lens}}\approx 0.53$) on the scales relevant for primordial $B$-mode searches ($30 \leq l \leq 300$). This is the highest delensing efficiency achieved on real $B$-modes to date. Considering lensing $B$-modes to be approximately equivalent to $5\,\mu\text{K-arcmin}$ white noise on large scales, we find that this level of delensing could reduce $\sigma(r)$ by a factor 1.3 for an ideal survey with a uniform map depth of $5\,\mu\text{K-arcmin}$, and by a factor 1.7 for SO's five-year goal depth of $2\,\mu\text{K-arcmin}$. Note that these estimates reflect the maximum $\sigma(r)$ improvement achievable in an idealized scenario; in reality, foregrounds, noise inhomogeneities and partial sky coverage result in more modest gains.

In all cases investigated above, the observed effect of delensing was consistent with theoretical expectations and detected with high statistical significance ($19.7\sigma$ for ACT, $10\sigma$ and $4.3\sigma$ for Planck in the full and restricted multipole ranges, respectively). Our results appear robust to various systematic effects including uncertainties in the modelling of lensing tracers, survey non-idealities and Galactic foreground contamination.

The delensing methods demonstrated here are directly applicable to upcoming CMB datasets, including early observations now being collected by SO. Cross-correlating our template with new polarization maps could facilitate an initial characterization of lensing $B$-modes, even in the regime where the $BB$ power spectrum remains noise-dominated; this could serve as a useful tool to understand instrumental systematics such as filtering and transfer functions. The present work therefore constitutes an important step towards future delensing analyses and improved constraints on $r$ with SO and other forthcoming surveys.

Although the impact of delensing on $\sigma(r)$ will be modest in the earliest stages, the steadily increasing sensitivity of current and next-generation experiments will lead to more significant improvements over time. As this occurs it will be important to ensure that Galactic and extragalactic foregrounds, which affect the CIB~\cite{lizancos_delensing_2022} and the internal lensing reconstruction~\cite{beck_impact_2020, lizancos_impact_2022, baleatolizancosHaloModelExtragalactic2025}, respectively, remain under control. Residual contamination and subsequent biases to the delensed $B$-mode power spectrum will be quantified using realistic simulations as well as null tests involving different combinations of frequency channels and masking schemes.

Our template construction pipeline can also be naturally extended to include deeper $E$-mode observations and $\kappa$ reconstructions (for example from the SO LAT), alongside the ACT data used here. Further gains in delensing efficiency may be achieved by incorporating new galaxy samples such as spectroscopically-confirmed emission line galaxies (ELGs) from DESI~\cite{ross_construction_2024}, as well as future data from Euclid~\cite{ilic_euclid_2022} and Rubin LSST~\cite{lsst_science_collaboration_lsst_2009,ivezic_lsst_2018}. Direct harmonic analysis methods~\cite{lizancos_harmonic_2024} may then be implemented to refine the treatment of these discrete tracers and avoid pixelization issues.

Finally, while the present work focuses on $B$-mode delensing for PGW searches, our co-added $\kappa$ map is also suitable for a broader range of uses including temperature and $E$-mode delensing. This is especially important for bispectrum analyses aiming to constrain primordial non-Gaussianity~\cite{abazajian_cmb-s4_2016}. Drawing on the wealth of high-quality data provided by current and upcoming cosmological surveys, multi-tracer delensing techniques will therefore remain an important tool for early-Universe science in the next few years.

\begin{acknowledgments}
The authors would like to thank Gerrit Farren for providing the unWISE overdensity maps used in this work. We acknowledge the use of the following public software packages: \texttt{CAMB}~\cite{lewis_efficient_2000}, \texttt{healpy}~\cite{zonca_healpy_2019}, \texttt{cmblensplus}~\cite{namikawa_cmblensplus_2021}, \texttt{NaMaster}~\cite{alonso_unified_2019}, \texttt{NumPy}~\cite{harris_array_2020}, \texttt{SciPy}~\cite{virtanen_scipy_2020} and \texttt{matplotlib}~\cite{hunter_matplotlib_2007}. Numerical computations were carried out using resources at NERSC (National Energy Research Scientific Computing Center), a U.S. Department of Energy Office of Science User Facility operated under Contract No. DE-AC02-05CH11231. EH is supported by a Gates Cambridge Scholarship (grant OPP1144 from the Bill \& Melinda Gates Foundation). AC acknowledges support from the STFC (grant numbers ST/S000623/1 and ST/X006387/1). MM acknowledges support from NSF grants AST-2307727 and AST-2153201. NS acknowledges support from DOE award number DE-SC0025309. CS acknowledges support from the Agencia Nacional de Investigaci\'on y Desarrollo (ANID) through Basal project FB210003. BS acknowledges support from
the European Research Council (ERC) under the European Union’s Horizon 2020 research and innovation
programme (Grant agreement No. 851274). Support for ACT was through the U.S.~National Science Foundation through awards AST-0408698, AST-0965625, and AST-1440226 for the ACT project, as well as awards PHY-0355328, PHY-0855887 and PHY-1214379. Funding was also provided by Princeton University, the University of Pennsylvania, and a Canada Foundation for Innovation (CFI) award to UBC. ACT operated in the Parque Astron\'omico Atacama in northern Chile under the auspices of the Agencia Nacional de Investigaci\'on y Desarrollo (ANID). The development of multichroic detectors and lenses was supported by NASA grants NNX13AE56G and NNX14AB58G. Detector research at NIST was supported by the NIST Innovations in Measurement Science program. Computing for ACT was performed using the Princeton Research Computing resources at Princeton University, the National Energy Research Scientific Computing Center (NERSC), and the Niagara supercomputer at the SciNet HPC Consortium. SciNet is funded by the CFI under the auspices of Compute Canada, the Government of Ontario, the Ontario Research Fund–Research Excellence, and the University of Toronto. We thank the Republic of Chile for hosting ACT in the northern Atacama, and the local indigenous Licanantay communities whom we follow in observing and learning from the night sky.
\end{acknowledgments}


\appendix

\section{Inclusion of DESI LRGs}\label{appendix_C}

In this appendix, we investigate whether a higher delensing efficiency can be achieved by including additional galaxy-overdensity maps in our analysis. We use data from the DESI Legacy Imaging Surveys (LS) DR9~\cite{dey_overview_2019}, which combine images of around $20,000\,\text{deg}^2$ of the Northern sky captured in three optical and four mid-infrared bands. Specifically, we consider the main sample of luminous red galaxies (LRGs) characterized in~\cite{zhou_desi_2023}, which was used to select targets for DESI~\cite{zhou_target_2023}.

\begin{figure}[h]
    \centering
    \includegraphics[width=\linewidth, height=0.25\textheight]{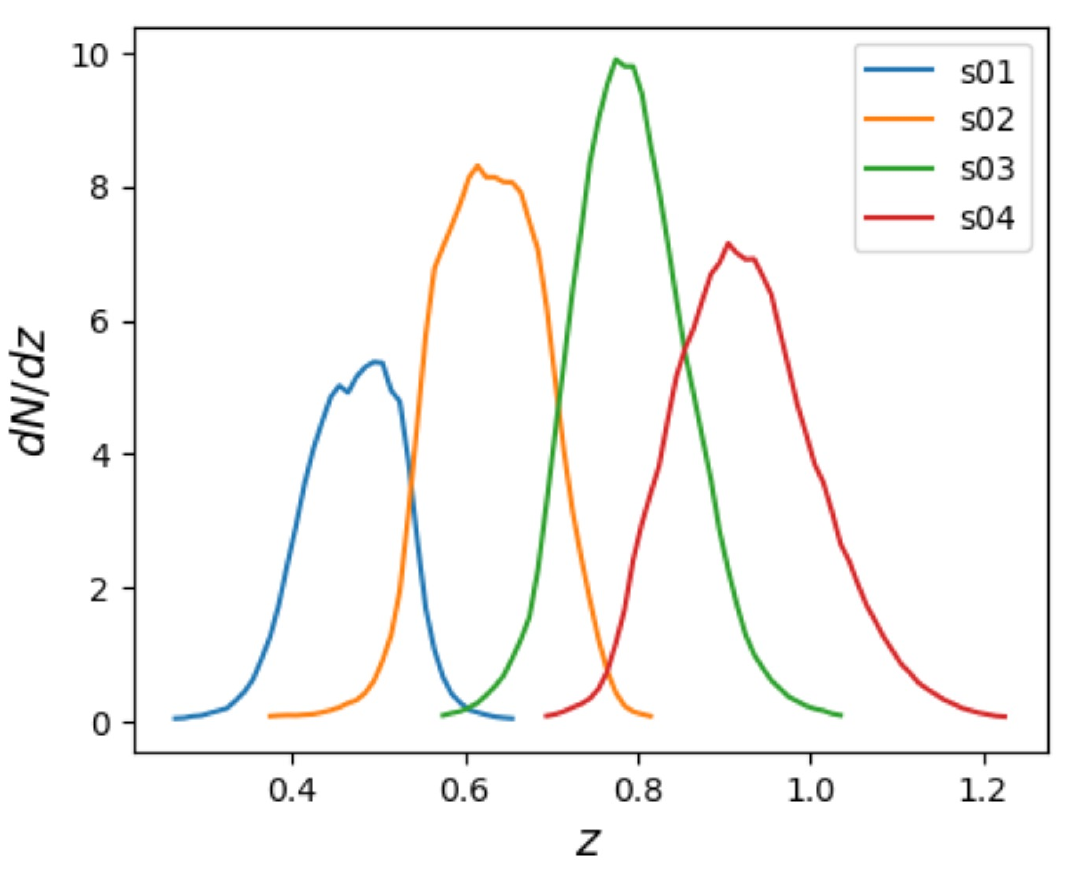}
    \vspace{-15pt}
    \caption{Normalized redshift distributions of the four DESI LRG galaxy samples.}
    \label{fig:desi_dNdz}
\end{figure}

This LRG sample is divided into four tomographic bins whose redshift distributions, shown in Fig.~\ref{fig:desi_dNdz}, are measured using DESI spectroscopy. The survey footprint and additional masks applied to remove bright sources are described in~\cite{zhou_desi_2023}. The DESI LRG subsamples are highly uniform and exhibit low stellar contamination in the remaining area. These properties, in addition to the accurate knowledge of their redshift distributions, make them excellent candidates for cross-correlation analyses.

\begin{figure}[h]
    \centering
    \includegraphics[width=\linewidth, height=0.25\textheight]{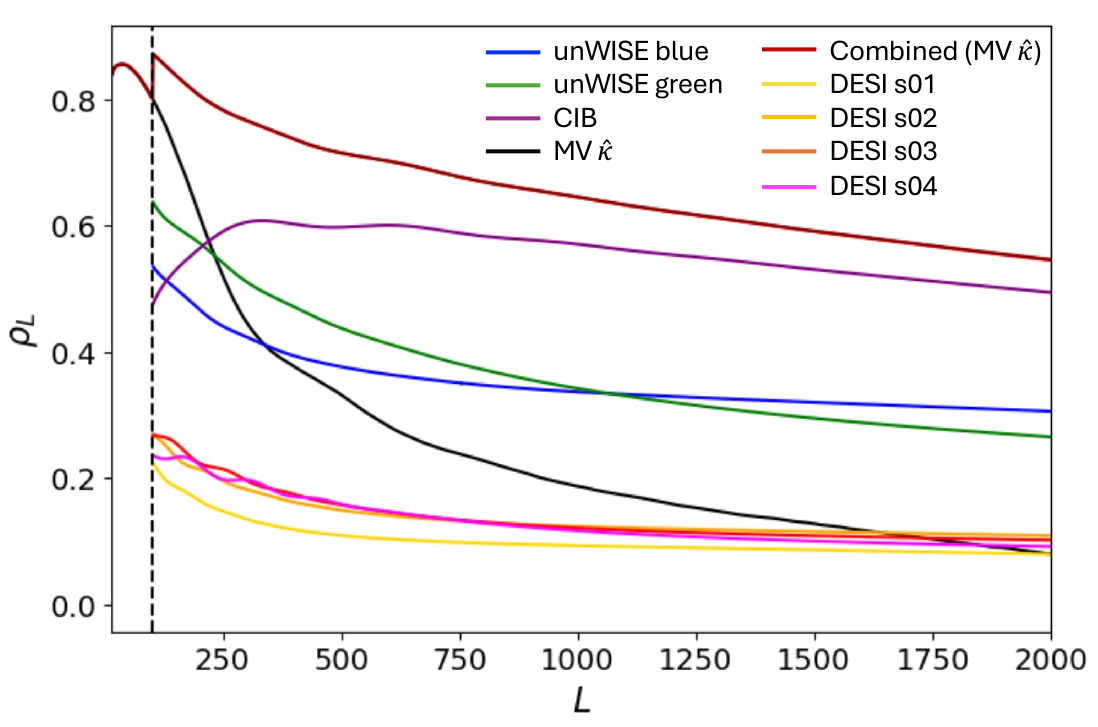}
    \vspace{-15pt}
    \caption{Same as Fig.~\ref{fig:rho_results}, with the addition of the four DESI LRG samples. These galaxy-overdensity maps have higher shot noise levels than unWISE and are thus significantly less correlated with the true $\kappa$. As a result, the difference in $\rho_L$ between the new combined tracer (dark red line) and the baseline case (bright red line) is indistinguishable on this plot.}
    \label{fig:rho_with_desi}
\end{figure}

Applying the steps outlined in Sec.~\ref{section:methodology} to model the cross-spectra between the DESI LRG samples and our other LSS tracers, we obtain the results shown in Fig.~\ref{fig:rho_with_desi}. The correlation coefficients with CMB lensing are significantly lower for the DESI LRGs than for unWISE, causing the former to be heavily downweighted in the optimal linear combination of tracers (Eq.~\ref{eq:kappa_comb}). Overall, the co-added lensing reconstruction is nearly identical to the baseline case in Fig.~\ref{fig:rho_results}, with $\rho_L$ values differing by less than $1\%$ in the multipole range plotted here.

The lack of significant impact is explained by the lower surface density and narrower redshift distribution of the DESI LRGs compared to unWISE, leading to increased shot noise and reduced overlap with the CMB lensing kernel. Correlation coefficients may be improved by using the extended photometric sample mentioned in~\cite{zhou_desi_2023}, which has a higher number density of galaxies, and by weighting or redefining tomographic bins to maximize overlap with $\kappa$. It will also be important to account for possible systematics related to anisotropies in the redshift distributions~\cite{baleato_lizancos_impact_2023}; we leave a detailed exploration of these ideas for future work.

\section{ACT transfer functions}\label{appendix_A}

In the results presented in the main text, we multiply the measured $C_l^{\hat{B}B_t}$ and $C_l^{B_tB_t}$ by an $l$-wise transfer function to correct for the Fourier-space filter applied to the ACT CMB maps. This assumes the filtering operation is isotropic, which is a reasonable approximation at high $l$ but may not be as accurate on large scales. We now investigate the impact of transfer function anisotropy on the template power spectrum and its cross-spectrum with observed $B$-modes.

Working in 2D Fourier space, we can express the ACT DR6 filtering operation as follows:
\begin{equation}\label{eq:dust_sed}
f(\bm{l})=
    \begin{cases}
     0 & \text{if } |l\cos\phi_{\bm{l}}|\leq l^f_x \text{ or } |l\sin\phi_{\bm{l}}|\leq l^f_y\\    
   1 & \text{otherwise },   
    \end{cases}
\end{equation}
where $\bm{l}=\left(l\cos\phi_{\bm{l}},l\sin\phi_{\bm{l}}\right)^T$ represents the wavevector, $l^f_x=90$ and $l^f_y=50$. For an isotropic CMB field $X(\bm{l})$, the power spectrum $C_l^{XX}$ is defined by
\begin{equation}\label{eq:ps_isotropic}
\langle X(\bm{l})X^*(\bm{l}')\rangle=(2\pi)^2\delta(\bm{l}-\bm{l}')C_l^{XX}.
\end{equation}
Including the anisotropic filter $f(\bm{l})$ causes the left-hand side to depend on the direction of $\bm{l}$; we can then compute an analogue of the power spectrum in Eq.~\eqref{eq:ps_isotropic} by averaging over $\phi_{\bm{l}}$, obtaining
\begin{equation}\label{eq:ps_aniso}
    \int_0^{2\pi}\frac{d\phi_{\bm{l}}}{2\pi}\langle f(\bm{l})X(\bm{l})f^*(\bm{l}')X^*(\bm{l}')\rangle=(2\pi)^2\delta(\bm{l}-\bm{l}')C_l^{XX,f}.
\end{equation}
Evaluating Eq.~\eqref{eq:ps_aniso} explicitly and noticing that $|f(\bm{l})|^2=f(\bm{l})$ leads to the relation
\begin{equation}\label{eq:tf_average}
    C_l^{XX,f}=\left(\int_0^{2\pi}\frac{d\phi_{\bm{l}}}{2\pi}f(\bm{l})\right)C_l^{XX}=\bar{f}_lC_l^{XX},
\end{equation}
where the isotropic transfer function $\bar{f}_l$ is given by~\cite{rosenberg_filtering_2024}
\begin{equation}\label{eq:tf_explicit}
    \bar{f}_l=1-\frac{2}{\pi}\left[\arcsin\left(\frac{l^f_x}{l}\right)+\arcsin\left(\frac{l^f_y}{l}\right)\right].
\end{equation}
The transfer functions estimated from simulations for $C_l^{EE}$ and $C_l^{BB}$ (see Sec.~\ref{section:methodology}) are in excellent agreement with this result.

We now propagate the impact of anisotropic filtering to power spectra involving the lensing template. The flat-sky analogue of Eq.~\eqref{eq:lensing_B} is
\begin{equation}\label{eq:lensing_B_flat}
    B^{\rm{lens}}(\bm{l})=\int\frac{d^2\bm{l'}}{(2\pi)^2}\mathcal{W}(\bm{l},\bm{l}')E(\bm{l}')\kappa(\bm{l}-\bm{l}'),
\end{equation}
where the mode-coupling function corresponds to~\cite{lewis_weak_2006}
\begin{equation}\label{eq:flat_W}
    \mathcal{W}(\bm{l},\bm{l}')=2\frac{\bm{l}'\cdot\left(\bm{l}-\bm{l}'\right)}{|\bm{l}-\bm{l}'|^2}\sin\left[2\left(\phi_{\bm{l}}-\phi_{\bm{l}'}\right)\right].
\end{equation}
The fields $E(\bm{l}')$ and $\kappa(\bm{l}-\bm{l}')$ are replaced by their Wiener-filtered counterparts to build the lensing template. Here, we consider a noise-free case for simplicity (so the Wiener filters reduce to unity) and to isolate the effect of $f(\bm{l})$; the template auto-spectrum then becomes
\begin{equation}\label{eq:temp_auto_aniso1}
    \int_0^{2\pi}\frac{d\phi_{\bm{l}}}{2\pi}\langle B^f_t(\bm{l})B_t^{f*}(\bm{l}')\rangle=(2\pi)^2\delta(\bm{l}-\bm{l}')C_l^{B_tB_t,f}
\end{equation}
with
\begin{multline}\label{eq:temp_auto_aniso2}
    \langle B^f_t(\bm{l})B_t^{f*}(\bm{l}')\rangle
    =(2\pi)^2\delta(\bm{l}-\bm{l}') \\ \times \int \frac{d^2\bm{l}''}{(2\pi)^2}|\mathcal{W}(\bm{l},\bm{l}'')|^2f(\bm{l}'')C_{l''}^{EE}C_{|\bm{l}-\bm{l}''|}^{\kappa\kappa}.
\end{multline}

\begin{figure}[h]
    \centering
    \includegraphics[width=\linewidth, height=0.25\textheight]{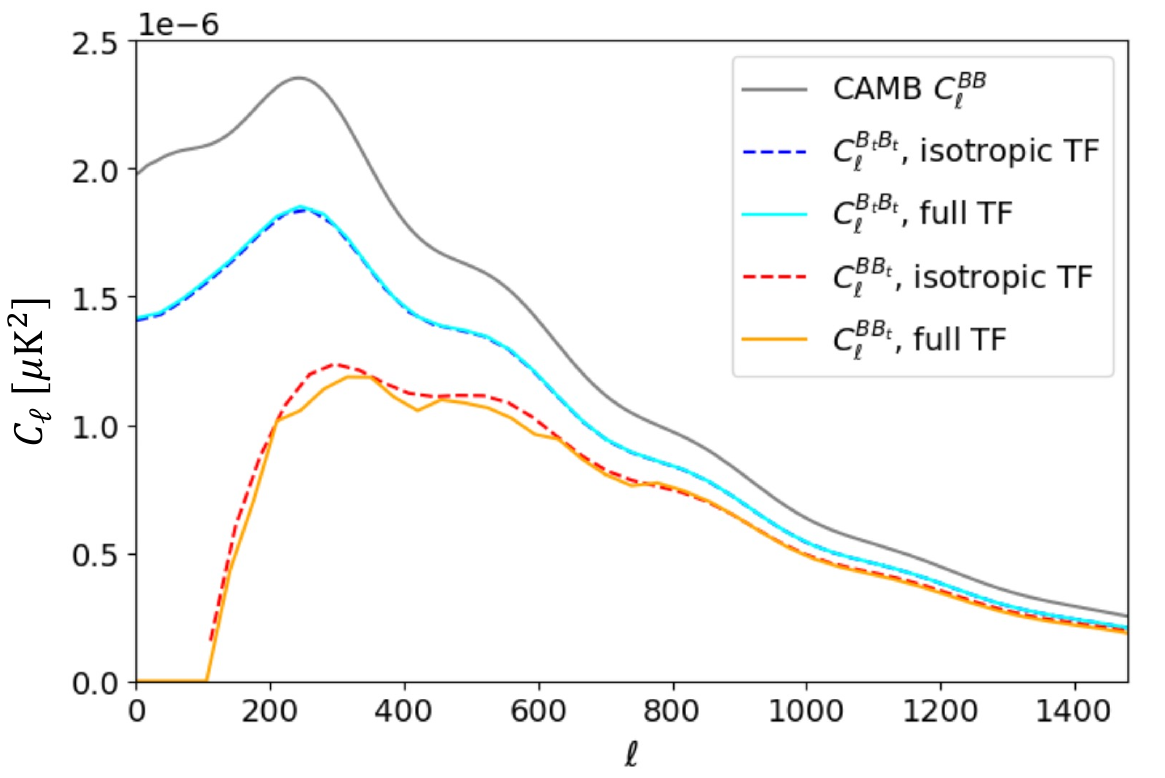}
    \vspace{-5pt}
    \caption{Impact of $k$-space filter anisotropy on the lensing template auto-spectrum and its cross-spectrum with observed $B$-modes in an idealized noise-free case. Without any filtering, $C_l^{B_tB_t}$ would approach the fiducial $C_l^{BB}$ shown in grey. The cyan line corresponds to the auto-spectrum of a template built with filtered $E$-modes, accounting for transfer function anisotropy (Eqs~\ref{eq:temp_auto_aniso1} and~\ref{eq:temp_auto_aniso2}). The yellow line shows the cross-spectrum with filtered $B$-modes computed in a similar way (Eqs~\ref{eq:temp_cross_aniso1} and~\ref{eq:temp_cross_aniso2}). These results are well approximated by the isotropic expressions in Eqs~\eqref{eq:temp_auto_iso} and~\eqref{eq:temp_cross_iso}, which are represented by the dashed blue and red lines, respectively.}
    \label{fig:transfer_functions}
\end{figure}

The cyan line in Fig.~\ref{fig:transfer_functions} is obtained by integrating Eqs~\eqref{eq:temp_auto_aniso1} and~\eqref{eq:temp_auto_aniso2} numerically for our fiducial cosmology. The excellent agreement with the isotropic approximation
\begin{equation}\label{eq:temp_auto_iso}
    C_l^{B_tB_t,\bar{f}}=\int \frac{d^2\bm{l}''}{(2\pi)^2}|\mathcal{W}(\bm{l},\bm{l}'')|^2\bar{f}_{l''}C_{l''}^{EE}C_{|\bm{l}-\bm{l}''|}^{\kappa\kappa},
\end{equation}
represented by the dashed blue line, is consistent with expectations: indeed, the anisotropy of $f(\bm{l})$ becomes negligible on the intermediate and small scales that contribute the most to lensing $B$-modes. 

We then compute the cross-spectrum between the lensing template and the filtered $B$-modes, which we define as follows:
\begin{equation}\label{eq:temp_cross_aniso1}
    \int_0^{2\pi}\frac{d\phi_{\bm{l}}}{2\pi}\langle f(\bm{l})B(\bm{l})B_t^{f*}(\bm{l}')\rangle=(2\pi)^2\delta(\bm{l}-\bm{l}')C_l^{BB_t,f}.
\end{equation}
Here, the ensemble average is given by
\begin{multline}\label{eq:temp_cross_aniso2}
    \langle f(\bm{l})B(\bm{l})B_t^{f*}(\bm{l'})\rangle = (2\pi)^2\delta(\bm{l}-\bm{l}')\\
    \times f(\bm{l})\int \frac{d^2\bm{l}''}{(2\pi)^2}|\mathcal{W}(\bm{l},\bm{l}'')|^2f(\bm{l}'')C_{l''}^{EE}C_{|\bm{l}-\bm{l}''|}^{\kappa\kappa},
\end{multline}
and the result of Eq.~\eqref{eq:temp_cross_aniso1} is shown as the yellow line in Fig.~\ref{fig:transfer_functions}. Small deviations from the isotropic case (dashed red line),
\begin{equation}\label{eq:temp_cross_iso}
    C_l^{BB_t,\bar{f}}=\bar{f}_{l}\int \frac{d^2\bm{l}''}{(2\pi)^2}|\mathcal{W}(\bm{l},\bm{l}'')|^2\bar{f}_{l''}C_{l''}^{EE}C_{|\bm{l}-\bm{l}''|}^{\kappa\kappa},
\end{equation}
appear at $l\leq 600$ but remain insignificant within the statistical uncertainties of our measurements.

Note that instrumental noise and survey masks are neglected in the idealized computations presented here. While Wiener filters do not break isotropy, incomplete sky coverage may interact in a non-trivial way with the Fourier-space transfer function. However, such effects were included in the simulations used to validate our pipeline and the resulting average power spectra did not significantly deviate from analytical predictions. We therefore conclude that our argument still holds and that $k$-space filtering anisotropy only has minimal impact on the power spectra of the lensing template and observed $B$-modes. This justifies our use of isotropic transfer function corrections at the end of Sec.~\ref{section:methodology}.

\section{Internal delensing bias}\label{appendix_B}

As explained in Refs~\cite{namikawa_cmb_2017} and \cite{lizancos_impact_2021}, including the $EB$ QE in the internal lensing reconstruction can lead to a suppression of the delensed $B$-mode power spectrum beyond the simple expectation in Eq.~\eqref{eq:delensed_B}.
The bias arises when the multipole range of the $B$-modes entering the QE overlaps with that of the $B$-modes to delens. For this reason, we removed QEs involving polarization fields when demonstrating delensing at intermediate and small scales (Fig.~\ref{fig:del_performance_ACT}).

\begin{figure}[h]
    \centering
    \includegraphics[width=\linewidth, height=0.25\textheight]{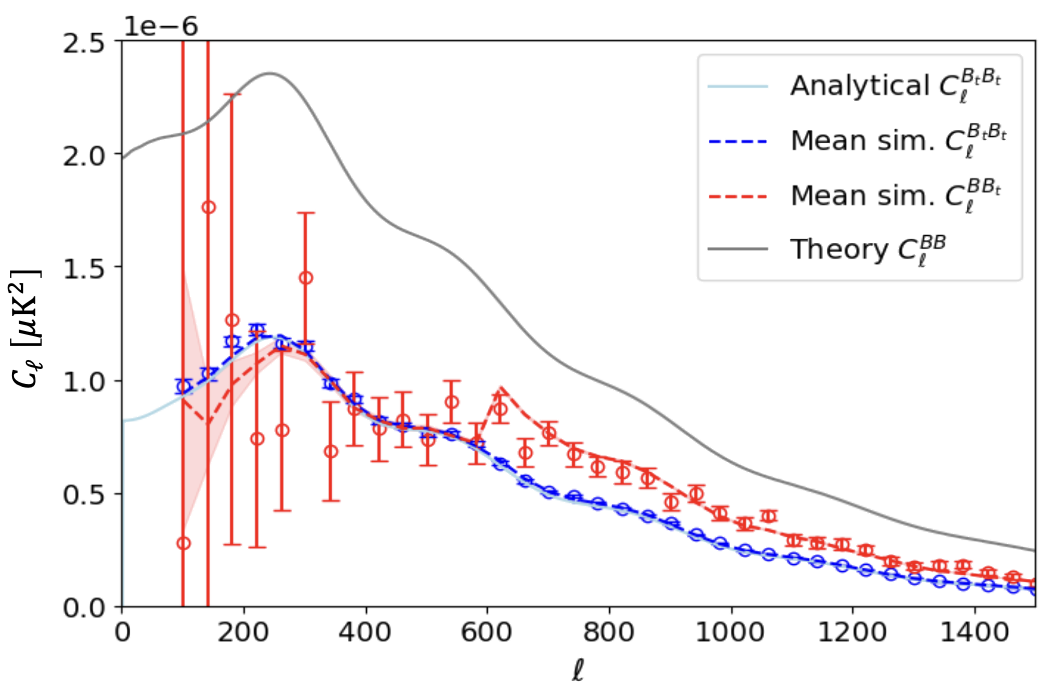}
    \vspace{-5pt}
    \caption{Same as Fig.~\ref{fig:temp_cross_auto}, but including the MV ACT DR6 lensing map in our multitracer template instead of its $\Theta$-only counterpart. Correlated noise realizations are used in the ACT-like CMB fields entering the template and in the simulated $B$-modes to delens. The internal bias on $C_l^{\hat{B}B_t}$ described in~\cite{lizancos_impact_2021} is clearly visible at $l\geq 600$, both in the measurements and average of simulations.}
    \label{fig:internal_bias}
\end{figure}

The red points in Fig.~\ref{fig:internal_bias} show the cross-spectrum of our template with ACT DR6 $B$-modes when the full MV combination of QEs is used instead. In the absence of bias (see Sec.~\ref{sec:results_tests}), this result should be consistent with the template auto-spectrum (blue points) and the analytical model (light blue line). Here, a clear excess of power appears at $l\geq 600$, which exactly corresponds to the minimum multipole of the CMB fields used to reconstruct the ACT lensing map. This positive bias $\Delta C_l^{\hat{B}B_t}$ comes from the disconnected part of the four-point function $\langle\hat{B}\times\hat{E}(\hat{E}\hat{B})\rangle$; assuming all fields are observed by a single telescope and the lensing reconstruction is purely based on the $EB$ QE, we obtain $\Delta C_l^{\hat{B}B_t}=D_lC_l^{\hat{B}\hat{B}}$ where $0<D_l\sim C_l^{BB,\rm{del}}/C_l^{\hat{B}\hat{B}}<1$~\cite{lizancos_impact_2021}. For current experiments, the main contribution to this term stems from the noise power spectrum $N_l^{BB}\gg C_l^{\tilde{B}\tilde{B}}$. 

We can therefore partially mitigate $\Delta C_l^{\hat{B}B_t}$ by extracting the $B$-modes to be delensed and those used for template construction from different instruments, as will be done for SO with the SATs and LAT. Combining the $EB$ reconstruction with other QEs and with external mass tracers further reduces the remaining bias terms. Our results are consistent with these expectations: we do not observe a statistically significant bias when delensing Planck $B$-modes with a multitracer template involving ACT QEs (Fig.~\ref{fig:del_performance_Planck}), and the excess power in Fig.~\ref{fig:internal_bias} is reproduced with simulations (dashed red line) when using correlated noise realizations at the $\kappa$ reconstruction and delensing stages (but disappears if the noise maps are independent). While bias terms involving the $B$-mode signal are clearly subdominant here, they will need to be mitigated at the sensitivity levels of SO and Stage-IV surveys.

The template auto-spectrum contains biases coming from disconnected parts of the six-point function $\langle\hat{E}(\hat{E}\hat{B})\times \hat{E}(\hat{E}\hat{B})\rangle$, given by $\Delta C_l^{B_tB_t}=D_l^2C_l^{\hat{B}\hat{B}}+D_lC_l^{W}$~\cite{lizancos_impact_2021}. Here, $C_l^{W}$ is the unbiased template power spectrum determined by the Wiener filters in Eq.~\eqref{eq:Cl_tempxtemp}, which is small compared to the noise-dominated $C_l^{\hat{B}\hat{B}}$. The first term is also significantly smaller than $\Delta C_l^{\hat{B}B_t}$ as $D_l\ll 1$ at current sensitivity levels. This explains why our measurements of the template auto-spectrum (blue points in Fig.~\ref{fig:internal_bias}) are consistent with theory within statistical uncertainties.

Overall, the bias on the power spectrum of delensed $B$-modes $C_l^{\hat{B}\hat{B},\rm{del}}=C_l^{\hat{B}\hat{B}}-2C_l^{\hat{B}B_t}+C_l^{B_tB_t}$ is dominated by the negative contribution of $-2\Delta C_l^{\hat{B}B_t}$; this additional suppression can bring residuals close to zero and deceptively mimic near-perfect delensing. While the effect can be avoided by removing the $EB$ QE from the lensing reconstruction (as was done in the present work when delensing ACT $B$-modes), this leads to a non-negligible loss of delensing efficiency and will degrade results more significantly in the future. For upcoming PGW searches requiring powerful delensing, the most advantageous way of mitigating internal biases is to restrict the $EB$ QE to a multipole range that does not overlap with the scales of interest for inflation ($l \leq 300$)~\cite{lizancos_impact_2021}.

\bibliography{apssamp}

\end{document}